\documentclass[runningheads]{svmult}
\input psfig
\usepackage{makeidx}   
\usepackage{graphicx}  
\usepackage{subeqnar}  
\usepackage{multicol}  
\usepackage{physprbb}  
\makeindex             

\begin{document}

\title*{An introduction to real-time renormalization group}
\toctitle{An introduction to real-time renormalization group}
\titlerunning{An introduction to real-time renormalization group}

\author{Herbert Schoeller \inst{1,2}}
\authorrunning{Herbert Schoeller}
\institute{Forschungszentrum Karlsruhe, Institut f\"ur
Nanotechnologie, 76021 Karlsruhe, Germany
\and
Institut f\"ur Theoretische Festk\"orperphysik, Universit\"at
Karlsruhe, 76128 Karlsruhe, Germany}

\maketitle              

\section{Introduction}\label{Introduction}
This article presents a tutorial introduction to a recently 
developed real-time renorma\-lization group method 
\cite{RG-anderson}. It describes nonequilibrium 
properties of discrete quantum systems coupled linearly to an 
environment. We illustrate the technique by a simple and
exactly solvable model: A quantum dot consisting of a single 
non-degenerate level coupled to two reservoirs. The article 
is intended for advanced students. Besides elementary 
quantum mechanics and statistical mechanics, it requires 
knowledge of second quantization and Wick's theorem. 
The latter topics can be learned easily from standard textbooks, 
see e.g. Ref.~\cite{fet-val}.

Renormalization group (RG) methods are standard tools to 
describe various aspects of condensed matter problems 
beyond perturbation theory \cite{anderson-wilson}.
Many impurity problems have been treated by numerical RG 
with excellent results both for thermodynamic quantities 
and spectral densities \cite{nrg,costi}. These RG techniques, 
however, cannot describe nonequilibrium properties like
the nonlinear conductance, the nonequilibrium stationary state, 
or the full time development of an initially out-of-equilibrium 
state. To address these aspects we present here a perturbative 
RG method, formulated for strongly correlated quantum systems 
with a finite number of states coupled linearly to external 
heat or particle reservoirs. Examples are: spin boson models,  
molecules interacting with electrodynamic fields, generalized 
Anderson-impurity models, quantum dot devices, 
magnetic nanoparticles interacting with phonons, etc..
Fundamentally new, we generate 
non-Hamiltonian dynamics during RG, which captures the 
physics of finite life times and dissipation. 
Furthermore, no initial or final cutoff in energy or time 
space is needed, i.e., large and small energy scales are 
accounted for correctly like in flow-equation methods 
\cite{flow equation}. Although correlation functions 
can also be studied, physical quantities like spin and 
charge susceptibilities or the current can be calculated directly 
without the need of nonequilibrium Green's functions. 

The purpose of our RG technique is to describe quantum 
fluctuations which are induced by strong coupling between a 
small quantum system and an environment. There are several 
recent experiments which show the importance of quantum 
fluctuations in metallic single-electron transistors 
\cite{joy-etal} and semiconductor quantum dots 
\cite{kondo-exp} (see \cite{schoeller} 
for an overview over theoretical papers). Due to the 
renormalization of resistance and local energy excitations, 
anomalous line shapes of the conductance have been observed, 
which can not be explained by golden-rule theories. For 
applications of the real-time RG to these cases we refer to 
Rfs.~\cite{RG-anderson,seb}.

Here we want to illustrate the method by an exactly sovable 
model, namely a quantum dot with one non-degenerate state 
with energy $\epsilon$ coupled to two reservoirs ($r=L,R$). 
The Hamiltonian $H=H_R+H_0+H_T$ consists of
three parts, corresponding to the reservoirs, the dot, and
tunneling 
\begin{eqnarray}
  H_R &=& \sum_r H_r = \sum_{kr} \epsilon_{kr} 
  a_{kr}^\dagger a_{kr} \,\,,\label{Hres}\\
  H_0 &=& \epsilon \,\,c^\dagger c \,\,,\label{Hdot}\\
  H_T &=& \sum_{kr}\left\{T_r a^\dagger_{kr} c\,+
  \,T_r^* c^\dagger a_{kr}
\right\} \,\,.\label{Htun}
\end{eqnarray}
Nonequilibrium is taken into account by describing the electrons in 
the reservoirs by Fermi distribution functions with different 
electrochemical potentials $\mu_r$. Tunneling is switched on suddenly at
the initial time $t_0$, i.e. initially the density matrix 
$\rho(t_0)=\rho_0$ decouples into an equilibrium part 
for each reservoir, $\rho_r=Z_r^{-1} e^{-\beta (H_r-\mu_r N_r)}$, 
and an arbitrary initial distribution $p(t_0)=p_0$
for the dot
\begin{equation}
  \rho_0 = p_0 \rho_{res} = p_0 \rho_L \rho_{res}\,\,.
\end{equation}
The aim is to calculate the time evolution of the reduced 
density matrix of the dot, $p(t)=Tr_{res} \rho(t)$, and the tunneling 
current $\langle I_r\rangle(t)$ flowing from reservoir $r$ 
to the dot. The tunneling current operator is given by
\footnote{Throughout this work we set $\hbar=k_B=1$ and use $e<0$}
\begin{equation}\label{cur-op}
  I_r = -e\dot{N_r} = 
  ie\sum_{k}\left\{T_r a^\dagger_{kr} c\,-\,T_r^* c^\dagger a_{kr}
    \right\} \,\,.
\end{equation}

The solution of the above quadratic Hamiltonian is
trivial since all degrees of freedoms can easily be integrated out.
Doing this by using Wick's theorem for all field operators
within the Keldysh formalism, one can easily solve the
full nonequilibrium problem \cite{car-etal}. However, except for 
having solved a special and almost trivial problem, we would
not have gained anything for solving more general
problems of dissipative quantum mechanics. Usually,
the local system can not be integrated out due to interaction
terms or spin degrees of freedom. Therefore, we try to proceed
differently. We will only integrate out the reservoirs
and keep the dot degrees of freedom explicitly. This is always 
possible for an effectively noninteracting bath. As a
result, we get an effective theory in terms of the local degrees of
freedom, expressed by a formally exact kinetic 
equation for the reduced density matrix of the dot. For the 
special Hamiltonian (\ref{Hres})-(\ref{Htun}), we
solve this equation exactly. Furthermore,
we will also develop a renormalization group method to 
solve the kinetic equation. We show that the RG equations 
describe the same exact solution. The important point is that 
both steps, i.e. (a) setting up the kinetic equation, and 
(b) setting up the RG equations, are not specific to the above 
Hamiltonian but can be applied to any discrete quantum system 
coupled linearly to an environment. The only difference is that, 
for most problems, the resulting RG equations have to be solved 
numerically. In conclusion, the above Hamiltonian serves as a 
test example to illustrate the RG technique and to demonstrate 
that it is well-defined and useful.

\section{Diagrammatic language}
\label{Diagrammatic language}

\subsection{Diagrams on the Keldysh contour}
\label{Diagrams on the Keldysh contour}

We start by introducing some convenient notations. The index
$\mu=\eta r$ labels the possible tunneling processes 
between reservoirs and dot. $\eta=\pm$ indicates 
tunneling in/out, and $r=L,R$ specifies the reservoir.
We define the following reservoir and dot operators
\begin{eqnarray}
&&j_{-r} = \sum_k T_r a^\dagger_{kr} \,\,,\,\,
j_{+r} = \sum_k T_r^* a_{kr} \,\,,\,\,
\label{def-j}\\
&&g_{-r} = c \,\,,\,\, g_{+r} = c^\dagger \,\,.
\label{def-g}
\end{eqnarray}
We denote the two possible states of
the dot by $s=0,1$ with energies $E_0=0$ and 
$E_1=\epsilon$. The Hamiltonian 
(\ref{Hres})-(\ref{Htun}) becomes
\begin{eqnarray}
  H_R &=& \sum_r \epsilon_{kr} a_{kr}^\dagger a_{kr} \,\,,
  \label{Hres2}\\
  H_0 &=& \sum_s E_s |s\rangle\langle s| \,\,,
  \label{Hdot2}\\
  H_T &=& \sum_\mu :g_\mu j_\mu: =
          \sum_r \left\{g_{+r}j_{+r}+j_{-r}g_{-r}\right\}\,\,,
  \label{Htun2}
\end{eqnarray}
where the symbol $:\dots:$ denotes normal ordering of
Fermi field operators but without sign change when
two operators are interchanged. The current operator 
(\ref{cur-op}) for the left reservoir is given by
\begin{equation}\label{cur-op2}
I_L = \sum_\mu :i_\mu j_\mu: \,\,,\,\,
i_\mu = -ie\eta g_\mu \delta_{rL}\,\,,\,\,(\mu=\eta r)\,\,,
\end{equation}
and a corresponding equation for $I_R$.

The time evolution of an arbitrary observable $a$ 
follows from the von Neumann equation
\begin{eqnarray}
\langle a\rangle(t)&=&Tr \,a \rho(t) = 
Tr \,a \,e^{-iH(t-t_0)} \rho_0 e^{iH(t-t_0)}\label{evol1}\\
&=& Tr e^{iH(t-t_0)} \,a \,e^{-iH(t-t_0)} \rho_0 \,\,,
\label{evol2}
\end{eqnarray}
where, in the last step, we have used cyclic invariance
under the trace. To get a matrix element of the reduced density 
matrix of the dot, $p(t)_{ss'}=\langle a\rangle(t)$, we need
$a=|s'\rangle\langle s|$. For the current, we
take $a=I_L=\sum_\mu :i_\mu j_\mu:$. 

\begin{figure}
  \centerline{\psfig{figure=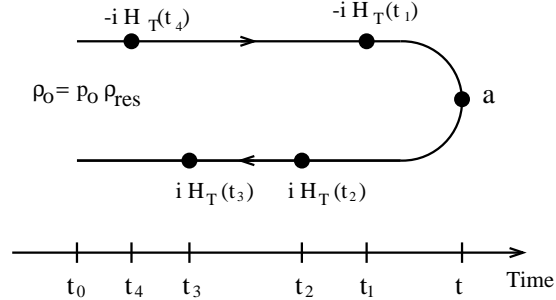,height=4cm}}
  \caption{Diagram corresponding to Eq.~(\ref{diagr}).}
\label{fig-contour}
\end{figure}

To integrate out the reservoirs, we expand the propagators 
in tunneling and apply Wick's theorem to the reservoir degrees
of freedom. We introduce the interaction picture 
\begin{equation}
b(t)=e^{i(H_R+H_0)(t-t_0)}\,b\, e^{-i(H_R+H_0)(t-t_0)}\,\,,
\end{equation}
and obtain
\begin{equation}
e^{iH(t-t_0)}\,a\,e^{-iH(t-t_0)} =
\tilde{T}\,e^{i\int_{t_0}^t dt'\,H_T(t')}\,a(t)\,
T\,e^{-i\int_{t_0}^t dt'\,H_T(t')}\,\,,
\label{T-eq}
\end{equation}
where $T$ and $\tilde{T}$ denote the time-ordering
and anti-time-ordering operators, respectively. Inserting 
(\ref{T-eq}) in (\ref{evol2}) and expanding in $H_T$ gives a
series of terms which we visualize diagrammatically,
see Fig.~\ref{fig-contour}. The upper (lower) line
corresponds to the forward (backward) propagator.
The diagram shown corresponds to the following
expression
\begin{equation}
i^2(-i)^2\,Tr\,H_T(t_3)H_T(t_2)\,a(t)\,H_T(t_1)H_T(t_4)
\,\rho_0\,\,.\label{diagr}
\end{equation} 
We see that the operators are ordered along a closed time 
path (Keldysh contour), as shown in Fig.~\ref{fig-contour}. 

The next step is to insert 
$H_T=\sum_\mu :g_\mu j_\mu:$ for the tunneling 
Hamiltonian and $\rho_0=p_0\rho_{res}$ for the initial 
density matrix. We use the short-hand
notation $g_i=g_{\mu_i}(t_i)$ and $j_i=j_{\mu_i}(t_i)$,
and decompose each diagram into a dot and reservoir part.
If $a=|s'><s|$ is a dot operator, we get from (\ref{diagr})
\begin{equation}
i^2(-i)^2\,\sum_{\mu_1\mu_2\mu_3\mu_4}\,
\{Tr_0\,g_3g_2\, a(t)\,g_1g_4\,p_o\}\,\,\,
\{Tr_{res}\,j_3j_2j_1j_4\,\rho_{res}\}
\,\,,
\label{example-dot}
\end{equation}
whereas, if $a=I_L=\sum_\mu :i_\mu j_\mu:$ is the current
operator, we get
\begin{equation}
i^2(-i)^2\,\sum_{\mu\mu_1\mu_2\mu_3\mu_4}\,
\{Tr_0\,g_3g_2\, i_\mu(t)\,g_1g_4\,p_o\}\,\,\,
\{Tr_{res}\,j_3j_2\,j_\mu(t)\,j_1j_4\,\rho_{res}\}
\,\,.
\label{example-cur}
\end{equation}
Here, $Tr_0$ ($Tr_{res}$) denotes the trace over the dot (reservoir)
degrees of freedom.
The reader can convince himself very easily that this
factorization does not imply any additional minus signs from 
commutation of Fermi operators.
The reason is the quadratic form of the tunneling Hamiltonian.

The trace over the reservoirs can be calculated easily
by using Wick's theorem \cite{fet-val}. As a result, we can 
decompose any average over products of reservoir field operators into
a sum over products of pair contractions. Denoting by 
$\langle\dots\rangle$ the average over the reservoirs, we
get for the reservoir part of (\ref{example-dot}) 
\begin{eqnarray}
  \langle j_3j_2j_1j_4 \rangle &=&
  j_3j_2j_1j_4 
  \begin{picture}(-32,11)
    \put(-32,8){\line(0,1){3}}
    \put(-32,11){\line(1,0){9}}
    \put(-23,8){\line(0,1){3}}
    \put(-15,8){\line(0,1){3}}
    \put(-15,11){\line(1,0){9}}
    \put(-6,8){\line(0,1){3}}
  \end{picture}
  \begin{picture}(32,11)
  \end{picture}
+ j_3j_2j_1j_4 
  \begin{picture}(-32,11)
    \put(-32,8){\line(0,1){3}}
    \put(-32,11){\line(1,0){17}}
    \put(-15,8){\line(0,1){3}}
    \put(-23,8){\line(0,1){6}}
    \put(-23,14){\line(1,0){17}}
    \put(-6,8){\line(0,1){6}}
  \end{picture}
  \begin{picture}(32,11)
  \end{picture}
+ j_3j_2j_1j_4 
  \begin{picture}(-32,11)
    \put(-32,8){\line(0,1){6}}
    \put(-32,14){\line(1,0){26}}
    \put(-6,8){\line(0,1){6}}
    \put(-23,8){\line(0,1){3}}
    \put(-23,11){\line(1,0){8}}
    \put(-15,8){\line(0,1){3}}
  \end{picture}
  \begin{picture}(32,11)
  \end{picture}
\nonumber \\
&=& \langle j_3j_2 \rangle\langle j_1j_4 \rangle -
\langle j_3j_1 \rangle\langle j_2j_4 \rangle +
\langle j_3j_4 \rangle\langle j_2j_1 \rangle 
\,\,.
\end{eqnarray}
Each pair contraction corresponds to an equilibrium 
average over two reservoir field operators. If two
contractions intersect proper minus signs have to
be taken into account due to Fermi statistics.
A pair contraction $\langle j_{\mu}(t) j_{\mu'}(t') \rangle$
depends on the relative time $t-t'$ and is only
non-zero for $\mu'=\bar{\mu}$, with 
$\bar{\mu}= -\eta r$ (if $\mu=\eta r$).
We define 
\begin{equation}
\gamma_\mu(t) = \langle j_{\bar{\mu}}(t) j_\mu \rangle\,\,,
\label{gamma-def}
\end{equation}
and get, by using the definition (\ref{def-j})
\begin{eqnarray}
\gamma_\mu(t) &=& \sum_k |T_r|^2\,
\left\{
\begin{array}{l@{\quad \mbox{for} \quad}r}
\langle a^\dagger_{kr}(t) a_{kr} \rangle & \eta=+ \\
\langle a_{kr}(t) a^\dagger_{kr} \rangle & \eta=- 
\end{array} \right. 
\nonumber\\
&=& {1\over 2\pi} \int dE \,\Gamma_r(E) e^{i\eta Et}
f^\eta(E-\mu_r) 
\label{gamma-exact}\\
&\cong& {\Gamma_r\over 2\pi} \int dE \,e^{\eta(E-\mu_r)/D} 
e^{i\eta Et} f^\eta_r(E) \,\,.
\label{gamma-zw}
\end{eqnarray}
Here we have defined 
\begin{equation}
\Gamma_r \cong \Gamma_r(E)=2\pi \sum_k |T_r|^2 
\delta(E-\epsilon_{kr})= 2\pi |T_r|^2 N_r(E)\,\,,
\label{Gamma-def}
\end{equation}
with $N_r(E)$ being the density of states of reservoir $r$.
We define $f^+=f$, $f^-=1-f$, and 
$f^\eta_r(E)=f^\eta(E-\mu_r)$, with 
$f(E)=1/(\exp(\beta E)+1)$ being the Fermi function,
and $\beta=1/T$ the inverse temperature.
We take the density of states $N_r(E)$ independent of energy
and regularize the integral (\ref{gamma-exact})
by introducing an exponential 
convergence factor $e^{\eta (E-\mu_r)/D}$.
Here, $D$ corresponds to a high energy cutoff. In the end,
we will send $D\rightarrow\infty$. Performing the integral
(\ref{gamma-zw}) gives 
\begin{equation}
\gamma_\mu(t) = {-i\Gamma_r e^{i\eta\mu_r t}\over
2\beta\sinh[\pi(t-i/D)/\beta]}\,\,.
\label{gamma-result}
\end{equation}
Furthermore, we define
\begin{equation}
\gamma_r^\eta(t) = \gamma_\mu(\eta t)\,\,,
\label{gamma-eta-def}
\end{equation}
and note the important property
\begin{equation}
\lim_{D\rightarrow\infty}\left\{
\gamma^+_r(t) + \gamma^-_r(t) \right\} = 
\Gamma_r \delta(t)\,\,,
\label{gamma-delta}
\end{equation}
which follows directly from (\ref{gamma-zw}).

\begin{figure}
  \centerline{\psfig{figure=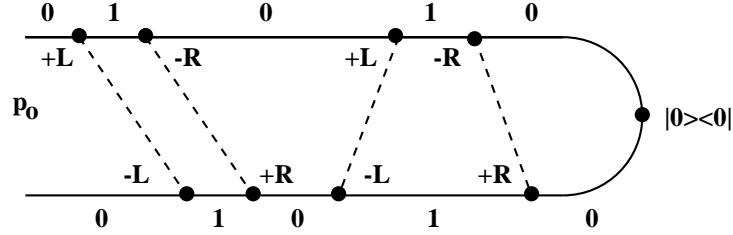,height=3cm}}
  \caption{Diagram after the application of Wick's theorem.
On the propagators the states $s=0,1$ of the dot are shown. At the
vertices we have indicated the index $\mu=\eta r$.}
\label{fig-diagram}
\end{figure}
We indicate the pair contractions diagrammatically by
connecting the corresponding vertices by a dashed line,
see Fig.~\ref{fig-diagram}. The remaining part for the 
dot degrees of freedom 
still remains, see Eq.~(\ref{example-dot}). 
We calculate this part by inserting intermediate states
of the dot between the operators. These states are
indicated in Fig.~\ref{fig-diagram} between the 
tunneling vertices. 
The reservoirs are already integrated out, so the 
tunneling vertices correspond to the dot 
operators $g_{\mu_i}$
(with an additional factor $\mp i$ for a vertex on the
upper (lower) propagator). Between the tunneling vertices
we have the free time evolution of the dot, i.e. for
a propagation of state $s$ from $t_2$ to $t_1$, we
get an exponential factor $e^{-i E_s (t_1-t_2)}$. 

Our diagrammatic language provides an
effective description in terms of the dot degrees
of freedom. The presence of the reservoirs is 
reflected by the retarded coupling of the tunneling 
vertices by the free Green's function of the reservoirs.
In particular, this means that the forward and backward
propagator are no longer independent but are coupled
by reservoir lines. We will see in section 
\ref{General approach} that
this leads to rates in a kinetic equation for $p(t)$.

\subsection{Superoperator notation}
\label{Superoperator notation}

In this section we will replace the double-propagator
diagrams on the Keldysh contour by a convenient matrix notation. 
This provides a very compact and analytic way to express
diagrams by formulas.

\begin{figure}
  \centerline{\psfig{figure=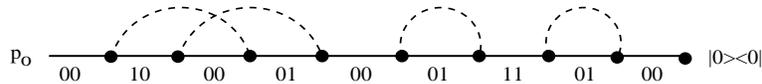,width=10cm}}
  \caption{Diagram in the double-state representation. It
results from Fig.~\ref{fig-diagram} by taking the upper and
lower line together to one single line.}
\label{fig-liouville}
\end{figure}
Instead of considering two propagators and specifying the dot
states on each propagator separately, we can formally take
both propagators together to one line and specify the 
state on this new propagator by a double dot state
$(s,s^\prime)$, see Fig.~\ref{fig-liouville}. Here, the
first (second) state corresponds to the state on the
upper (lower) propagator. By this trick we have lost 
the information wether a tunneling vertex lies on
the upper or lower propagator. To recover this we add to the
index $\mu$ of the tunneling vertex operator an
additional index $p=\pm$, which indicates wether $g_\mu$ acts
on the upper (lower) propagator. The new vertex is 
denoted by $G^p_\mu$ with matrix elements
\begin{eqnarray}
(G^+_\mu)_{s_1 s_1^\prime,s_2 s_2^\prime} &=& 
-i(g_\mu)_{s_1 s_2}\delta_{s_1^\prime s_2^\prime} 
\label{def-gmatrix1}\\
(G^-_\mu)_{s_1 s_1^\prime,s_2 s_2^\prime} &=& 
 i\delta_{s_1 s_2} (g_\mu)_{s_2^\prime s_1^\prime} \,\,.
\label{def-gmatrix2}
\end{eqnarray}
We have included the factors $\mp i$ for a vertex on the
upper (lower) propagator into the definition of $G^p_\mu$.
The free time evolution between
the vertices is given by a factor $e^{-i(E_s-E_{s'})(t_1-t_2)}$,
where $(s,s')$ indicates the double state on the line.
This can be written in operator form as 
$(e^{-iL_0(t_1-t_2)})_{ss',ss'}$, with
\begin{equation}
(L_0)_{s_1 s_1^\prime,s_2 s_2^\prime} =
\delta_{s_1 s_2}\delta_{s_1^\prime s_2^\prime}
(E_{s_1}-E_{s_1^\prime}) 
\,\,.\label{def-Lmatrix}
\end{equation}
Finally, a contraction connecting a vertex $G^{p'}_{\mu'}$ at
time $t'$ with a vertex $G^p_\mu$ at time $t>t'$, is denoted by 
$\gamma^{pp'}_{\mu\mu'}(t-t')$. Using the definition
(\ref{gamma-def}) and the fact that operators on the lower
propagator act always later than those on the upper propagator,
we obtain 
\begin{eqnarray}
\gamma^{pp'}_{\mu\mu'}(t) &=& \delta_{\bar{\mu},\mu'}
\left\{
\begin{array}{l@{\quad \mbox{for} \quad}r}
\langle j_\mu(t) j_{\bar{\mu}} \rangle = 
\gamma_{\bar{\mu}}(t) & p'=+ \\
\langle j_{\bar{\mu}} j_\mu(t) \rangle = 
\gamma_\mu(-t) & p'=-
\end{array} \right. 
\label{gamma-1}\\
&=& \delta_{\bar{\mu},\mu'}
{-i\Gamma_r e^{-i\eta \mu_r t}
\over 2 \beta\sinh[\pi(p't-i/D)/\beta]} \,\,,
\label{gamma-2}
\end{eqnarray}
where we used the result (\ref{gamma-result}) in the second step.

The double-state matrices 
$L_0$ and $G_\mu^p$ are called superoperators
in the sense that they act on single-state matrices, i.e. 
on ordinary operators. If $b$ is an ordinary operator, we
can define $L_0$ and $G_\mu^p$ by
\begin{equation}
L_0b=[H_0,b]\quad,\quad G^+_\mu b=-ig_\mu b
\quad,\quad G^-_\mu b=ibg_\mu\,\,.
\end{equation}

Within the superoperator notation and using 
$a=|s'\rangle\langle s|$, an arbitrary diagram for
$p_{ss'}(t)$ can be written as
\begin{equation}
\fbox{\parbox{8cm}{
\begin{center}
$p_{ss'}(t)\quad\rightarrow\quad\left\{e^{-iL_0(t-t_0)}\,
G_1G_2\dots\dots G_n
  \begin{picture}(-62,11)
    \put(-62,8){\line(0,1){3}}
    \put(-62,11){\line(1,0){25}}
    \put(-37,8){\line(0,1){3}}
    \put(-49,8){\line(0,1){6}}
    \put(-49,14){\line(1,0){21}}
    \put(-28,8){\line(0,1){6}}
    \put(-21,8){\line(0,1){3}}
    \put(-21,11){\line(1,0){12}}
    \put(-9,8){\line(0,1){3}}
  \end{picture}
  \begin{picture}(62,11)
  \end{picture}
\,p_0\right\}_{ss'}\,\,.$
\end{center}}}
\label{dia-liouville}
\end{equation}
The time dependence of $G_i= G_{\mu_i}^{p_i}(t_i)$ 
is defined by
\begin{equation}
G^p_\mu(t)=e^{iL_0(t-t_0)}G^p_\mu e^{-iL_0(t-t_0)}\,\,.
\label{G-intpic}
\end{equation}
All operators $G_1\dots G_n$ are coupled in all possible ways
by reservoir pair contractions, as indicated in (\ref{dia-liouville}).
Implicitly we assume summation over $\mu_1\dots\mu_n$
and $p_1\dots p_n$, together with the integration over the
time variables $t_1\dots t_n$ with $t>t_1>t_2>\dots >t_n>t_0$.

If $a=I_L=\sum_\mu :i_\mu j_\mu:$ corresponds to the current
operator, we get
\begin{equation}
\langle :i_\mu j_\mu:\rangle(t)
\quad\rightarrow\quad Tr_0 \,i_\mu\,\left\{e^{-iL_0(t-t_0)}\,
G_1G_2\dots\dots G_n
  \begin{picture}(-62,11)
    \put(-62,8){\line(0,1){3}}
    \put(-62,11){\line(1,0){25}}
    \put(-37,8){\line(0,1){3}}
    \put(-49,8){\line(0,1){6}}
    \put(-49,14){\line(1,0){21}}
    \put(-28,8){\line(0,1){6}}
    \put(-21,8){\line(0,1){3}}
    \put(-21,11){\line(1,0){12}}
    \put(-9,8){\line(0,1){3}}
  \end{picture}
  \begin{picture}(62,11)
  \end{picture}
  \begin{picture}(-131,11)
    \put(-131,8){\line(0,1){8}}
    \put(-131,16){\line(1,0){107}}
    \put(-24,8){\line(0,1){8}}
  \end{picture}
  \begin{picture}(131,11)
  \end{picture}
\,p_0\right\}\,\,.
\label{dia-liou-cur}
\end{equation}
In comparism to (\ref{dia-liouville}), we need an additional 
pair contraction to the current vertex. In order to treat the
boundary vertex $i_\mu$ as well within the superoperator
notation, we define a superoperator $I^p_\mu$ by
\begin{equation}
I^+_\mu b = i_\mu b/2\quad,\quad
I^-_\mu b = b\, i_\mu/2\,\,,
\end{equation}
with matrix elements
\begin{equation}
(I^+_\mu)_{s_1s_1^\prime,s_2s_2^\prime}=
{1\over 2}(i_\mu)_{s_1 s_2}\delta_{s_1^\prime s_2^\prime}
\quad,\quad
(I^-_\mu)_{s_1s_1^\prime,s_2s_2^\prime}=
{1\over 2}\delta_{s_1 s_2}(i_\mu)_{s_2^\prime s_1^\prime}
\,\,.
\label{def-curmatrix}
\end{equation}
Using cyclic invariance under the trace, 
we get for (\ref{dia-liou-cur}) 
\begin{equation}
\fbox{\parbox{8cm}{
\begin{center}
$\langle :i_\mu j_\mu:\rangle(t)
\quad\rightarrow\quad Tr_0 \,
I_t\,G_1G_2\dots\dots G_n
  \begin{picture}(-62,11)
    \put(-62,8){\line(0,1){3}}
    \put(-62,11){\line(1,0){25}}
    \put(-37,8){\line(0,1){3}}
    \put(-49,8){\line(0,1){6}}
    \put(-49,14){\line(1,0){21}}
    \put(-28,8){\line(0,1){6}}
    \put(-21,8){\line(0,1){3}}
    \put(-21,11){\line(1,0){12}}
    \put(-9,8){\line(0,1){3}}
  \end{picture}
  \begin{picture}(62,11)
  \end{picture}
  \begin{picture}(-72,11)
    \put(-72,9){\line(0,1){7}}
    \put(-72,16){\line(1,0){48}}
    \put(-24,9){\line(0,1){7}}
  \end{picture}
  \begin{picture}(72,11)
  \end{picture}
\,p_0\,\,,$
\end{center}}}
\label{dia-liou-cur2}
\end{equation}
where $I_t=I^p_\mu(t)$, and the interaction picture 
is defined by
\begin{equation}
I^p_\mu(t) = I^p_\mu \,e^{-iL_0(t-t_0)}\,\,.
\end{equation}

Eqs.~(\ref{dia-liouville}) and (\ref{dia-liou-cur2}) are
the central result of this section. They relate the reduced
density matrix of the dot and the average current to 
diagrammatic expressions in a very compact and analytic
way. It turns out that the usage of superoperators simplifies
the notation considerably. We will see in sections 
\ref{Kinetic equation} and \ref{Renormalization group}
that the derivation of kinetic equations and renormalization
group equations is very transparent in this language.
However, one should always keep in mind that the usage of
superoperators is only a formal trick to find a convenient
matrix notation. Therefore, we have set
up the diagrammatic representation in terms of the Keldysh
contour first, and then, in a second step, introduced the
superoperators. Of course it is also possible to start 
directly with superoperators \cite{RG-anderson},
which provides a more compact and shorter way to arrive
at Eqs.~(\ref{dia-liouville}) and (\ref{dia-liou-cur2}).
However, for pedagogical and physical reasons, we did not 
proceed in this way here.
The diagrams on the Keldysh contour reveal
better that there are different kinds of terms which have
to be distinguished very carefully from a physical point 
of view. Reservoir lines connecting the upper and lower
propagator correspond to rates, they change the state of
the dot simultaneously on the upper {\it and} the lower
propagator. This describes a transition from one diagonal 
matrix element of the reduced density matrix $p(t)$ to
another one. Such processes can not be expressed
on a Hamiltonian level and lead basically to the physics of
dissipation. In contrast, reservoir contractions which
connect vertices within the upper or lower propagator 
describe renormalization and broadening of levels.

\section{Kinetic equation}
\label{Kinetic equation}

\subsection{General approach}
\label{General approach}

In this section we will derive a self-consistent equation
for the reduced density matrix $p(t)$ of the dot,
together with an expression for the average current.
To achieve this it is essential to distinguish in 
Eqs.~(\ref{dia-liouville}) and (\ref{dia-liou-cur2}) between
connected and disconnected parts. From
(\ref{dia-liouville}) we see that any diagram for $p(t)$
can be written in the form
\begin{eqnarray}
e^{-iL_0(t-t_1)}\,\,(A_1G\dots GB_2)_{con}\,\,
e^{-iL_0(t_2-t_3)}\,\,(A_3G\dots GB_4)_{con}\dots
\nonumber\\
\dots e^{-iL_0(t_{2n-2}-t_{2n-1})}\,\,
(A_{2n-1}G\dots GB_{2n})_{con}\,\,e^{-iL_0(t_{2n}-t_0)}p_0\,\,.
\label{series}
\end{eqnarray}
Here, $(A_iG\dots GB_{i+1})_{con}$ denotes a sequence of 
vertices between $t_{i+1}$ and $t_i$ which are coupled by 
pair contractions in such a 
way that any vertical cut between $t_{i+1}$ and $t_i$ will 
cross some contraction. We define such a block as a 
connected part of a diagram. E.g. Fig.~\ref{fig-liouville}
shows a sequence of three connected blocks. 
The boundary vertices $A$ and
$B$ are identical to $G$, i.e. $A^p_\mu=B^p_\mu=G^p_\mu$, but
the interaction picture is defined differently
\begin{equation}
A^p_\mu(t) = A^p_\mu e^{-iL_0(t-t_0)} 
\quad,\quad
B^p_\mu(t) = e^{iL_0(t-t_0)} B^p_\mu\,\,.
\label{AB-intpic}
\end{equation}
The reason is that we want the connected part 
$(A_iG\dots GB_{i+1})_{con}$ of a diagram to depend only
on the relative time argument $t_i-t_{i+1}$. Furthermore,
we distinguish the boundary vertices $A^p_\mu$ and 
$B^p_\mu$ from $G^p_\mu$ since, within the
renormalization group procedure developed in 
section \ref{Renormalization group}, 
the boundary vertices renormalize differently. 

We define the sum over all connected diagrams between
$t'$ and $t$ by the kernel $\Sigma(t-t')$
\begin{equation}
\fbox{\parbox{8cm}{
\begin{center}
$\Sigma(t-t')\quad\rightarrow\quad
(A_t G_1G_2\dots G_{2n} B_{t'})_{con}\,\,.$
\end{center}}}
\label{sigma}
\end{equation}
We note the important property that the kernel
$\Sigma(t-t')$ is independent of the initial time $t_0$
since all exponential factors $e^{\pm iL_0 t_0}$ 
arising from the interaction picture cancel within
$\Sigma$. Thus, in order to calculate $\Sigma$, we
can set $t_0=0$ in the definition
of the interaction picture of $A$, $B$, and $G$,
see Eqs.~(\ref{G-intpic}) and (\ref{AB-intpic}).

Using the definition (\ref{sigma}) for $\Sigma$ 
in (\ref{series}), we obtain
\begin{eqnarray}
p(t)&=& e^{-iL_0(t-t_0)}p_0 \,+\,
\sum_{n=1}^\infty \int_{t_0}^t dt_1\int_{t_0}^{t_1} dt_2
\dots \int_{t_0}^{t_{2n-1}} dt_{2n}\nonumber\\
&&e^{-iL_0(t-t_1)}\,\Sigma(t_1-t_2)\,
e^{-iL_0(t_2-t_3)}\,\Sigma(t_3-t_4)\dots\nonumber\\
&&\dots e^{-iL_0(t_{2n-2}-t_{2n-1})}
\Sigma(t_{2n-1}-t_{2n})\,e^{-iL_0(t_{2n}-t_0)}p_0\,\,.
\end{eqnarray}
Differentiating with respect to time gives the
kinetic equation
\begin{equation}
\fbox{\parbox{8cm}{
\begin{center}
$\dot{p}(t) + iL_0 p(t) = \int_{t_0}^t dt'
\Sigma(t-t') p(t')\,\,.$
\end{center}}}
\label{kinetic-equation}
\end{equation}
Since the r.h.s. of this equation is a convolution in
time space, we can formally solve this equation 
in Laplace space. We define the Laplace transform by
\begin{equation}
\tilde{p}(z)=\int_{t_0}^\infty dt\,e^{izt}p(t)
\quad,\quad
\tilde{\Sigma}(z)=\int_0^\infty dt\,e^{izt}\Sigma(t)\,\,,
\label{laplace}
\end{equation}
and get from (\ref{kinetic-equation}) the solution 
\begin{equation}
\tilde{p}(z)={i\over z-L_0-i\tilde{\Sigma}(z)}\,\,p_0\,\,.
\label{p(z)-solution}
\end{equation}
The time dependence $p(t)$ follows by reversing the 
Laplace transform
\begin{equation}
p(t) = \lim_{\eta\rightarrow 0} {1\over 2\pi}
\int_{-\infty + i\eta}^{\infty + i\eta}
dz\,e^{-izt} \tilde{p}(z)
= \lim_{\eta\rightarrow 0}{1\over 2\pi}
\int\limits_{-\infty}^\infty
d\omega\,e^{-i\omega t} \tilde{p}(\omega+i\eta)
\,\,.
\label{p(t)-solution}
\end{equation}
We remark that $\tilde{p}(z)$, defined by 
(\ref{laplace}), is analytic in the
upper half plane since $p(t)$ will approach a
stationary value for $t\rightarrow\infty$. Thus,
within the integration region of (\ref{p(t)-solution})
the integrand is well-defined.
We see that the integral kernel
$\tilde{\Sigma}(z)$ is the central object which has
to be calculated. The full time evolution of $p(t)$ 
out of an arbitrary
nonequilibrium state can be obtained once 
$\tilde{\Sigma}(z)$ is known for all $z=\omega+i0^+$. 
The calculation of $\tilde{\Sigma}(z)$ will be the 
subject of the renormalization group approach described 
in section \ref{Renormalization group}. 

The stationary state is defined by
\begin{equation}
p_{st}=\lim_{t\rightarrow\infty}p(t)
=-i\lim_{z\rightarrow i0^+} z\,\tilde{p}(z)\,\,.
\end{equation}
Multiplying (\ref{p(z)-solution}) by 
$z[z-L_0-i\tilde{\Sigma}(z)]$ and taking the limit
$z\rightarrow i0^+$, we see that the stationary state
is the eigenvector of $L_0+i\tilde{\Sigma}(i0^+)$ with
eigenvalue zero
\begin{equation}
[L_0\,+\,\tilde{\Sigma}(i0^+)]\,p_{st} \,= \,0
\,\,.
\label{pst-solution}
\end{equation}

The density matrix $p(t)$ is hermitian. In Laplace
space this is equivalent to 
\begin{equation}
\tilde{p}(z)^*_{ss'}\,=\,\tilde{p}(-z^*)_{s's}\,\,.
\label{p(z)-symmetry}
\end{equation}
This follows from the solution
(\ref{p(z)-solution}) by using the symmetry relations
\begin{equation}
(iL_0)_{s_1s_1^\prime,s_2s_2^\prime}^*=
(iL_0)_{s_1^\prime s_1,s_2^\prime s_2}
\quad , \quad
(G^p_\mu)_{s_1s_1^\prime,s_2s_2^\prime}^*=
(G^{\bar{p}}_{\bar{\mu}})_{s_1^\prime s_1,s_2^\prime s_2}\,\,,
\label{LG-symmetry}
\end{equation}
where $\bar{p}=-p$. They 
follow directly from (\ref{def-gmatrix1}), (\ref{def-gmatrix2}), 
(\ref{def-Lmatrix}), and the hermiticity of the
Hamiltonian. The consequence for the kernel
(\ref{sigma}) is
\begin{equation}
\Sigma(t)_{s_1s_1',s_2s_2'}^*=\Sigma(t)_{s_1's_1,s_2's_2}
\quad,\quad 
\tilde{\Sigma}(z)_{s_1s_1',s_2s_2'}^*=
\tilde{\Sigma}(-z^*)_{s_1's_1,s_2's_2}\,\,.
\label{sigma-symmetry}
\end{equation}
Using these relations together with the hermiticity
of the initial density matrix $p_0$, we find directly
(\ref{p(z)-symmetry}) from (\ref{p(z)-solution}).

To prove conservation of probability
$Tr_0 p(t)=1$, we first note that
\begin{equation}
\sum_s (L_0)_{ss,\cdot\cdot}=  
\sum_{sp} (A^p_\mu)_{ss,\cdot\cdot}=  
\sum_{sp} (B^p_\mu)_{ss,\cdot\cdot}=  
\sum_{sp} (G^p_\mu)_{ss,\cdot\cdot}= 0\,\,.
\label{LABG-sum}
\end{equation}
Applying this to (\ref{sigma}) together with the fact 
that the contraction connected to the boundary operator
$A^p_\mu$ does not depend on $p$, we find the same property
for the kernel
\begin{equation}
\sum_s \Sigma(t)_{ss,\cdot\cdot} = 
\sum_s \tilde{\Sigma}(z)_{ss,\cdot\cdot} = 0\,\,.
\label{sigma-sum}
\end{equation}
Applying $Tr_0$ to the kinetic equation 
(\ref{kinetic-equation}), and using the properties
(\ref{LABG-sum}) and (\ref{sigma-sum}), we
find $d/dt Tr_0 p(t)=0$ which proves conservation
of probability.

To calculate the current we proceed analogously. From
(\ref{dia-liou-cur2}) and (\ref{series}) we see 
that any diagram for the average current
can be written as
\begin{equation}
Tr_0\,(I_t\,G\dots G B_{t'})_{con}\,p(t')\,\,.
\label{series-cur}
\end{equation}
The first connected block contains the current vertex.
The sum over all connected diagrams containing the
current vertex is denoted by $\Sigma_I(t-t')$
\begin{equation}
\fbox{\parbox{8cm}{
\begin{center}
$\Sigma_I(t-t')\quad\rightarrow\quad
(I_t G_1G_2\dots G_{2n} B_{t'})_{con}\,\,.$
\end{center}}}
\label{sigma-cur}
\end{equation}
The only difference to (\ref{sigma}) is that the
boundary vertex $A^p_\mu$ has been replaced by the
current vertex $I^p_\mu$. From (\ref{series-cur})
and (\ref{sigma-cur}) we find 
\begin{equation}
\fbox{\parbox{8cm}{
\begin{center}
$\langle I_L \rangle (t) = 
\sum_\mu \langle :i_\mu j_\mu: \rangle =
\int_{t_0}^t dt'\,Tr_0\,\Sigma_I(t-t')\,p(t')\,\,.$
\end{center}}}
\label{i(t)-solution}
\end{equation}
In Laplace space we get
\begin{equation}
\tilde{\langle I_L \rangle}(z) = 
Tr_0 \tilde{\Sigma}_I(z)\,\tilde{p}(z)\,\,,
\label{i(z)-solution}
\end{equation}
and the stationary solution follows from
\begin{equation}
\langle I_L \rangle_{st} = 
Tr_0 \,\tilde{\Sigma}_I(i0^+)\,p_{st}\,\,.
\end{equation}

The expectation value of the current is real, i.e.
$\tilde{\langle I_L \rangle}(z)^*=
\tilde{\langle I_L \rangle}(-z^*)$. This can be
seen from the solution (\ref{i(z)-solution}) by
using the symmetry relations
\begin{equation}
(I^p_\mu)_{s_1s_1^\prime,s_2s_2^\prime}^*=
(I^{\bar{p}}_{\bar{\mu}})_{s_1^\prime s_1,s_2^\prime s_2}\,\,,
\label{I-symmetry}
\end{equation}
and
\begin{equation}
\Sigma_I(t)_{s_1s_1',s_2s_2'}^*=\Sigma_I(t)_{s_1's_1,s_2's_2}
\quad,\quad 
\tilde{\Sigma}_I(z)_{s_1s_1',s_2s_2'}^*=
\tilde{\Sigma}_I(-z^*)_{s_1's_1,s_2's_2}\,\,.
\label{sigmaI-symmetry}
\end{equation}

\subsection{Exact solution}
\label{Exact solution}

For the model of a single non-degenerate
dot state, the kinetic equation and the current
formula can be solved exactly. We derive this
solution here and will check in section 
\ref{Exact solution of the RG equations}
that the renormalization group equations describe
the same solution. If the reader is not interested
in technical details, he can find the final results 
in Eqs.~(\ref{exact-relation}) and 
(\ref{exact-solution}), and can proceed to the
next section.

There are only
two possible dot states $s=0,1$. We denote by
$\bar{s}$ the conjugate state, $\bar{s}=1(0)$
if $s=0(1)$. Furthermore we define $\bar{p}=-p$
and $\bar{\mu}=-\eta r$ if $\mu=\eta r$. 
The following three properties are needed 
for the following
\begin{eqnarray}
\sum_{ps}(
G^p_\mu
  \begin{picture}(-8,11)
    \put(-8,8){\line(0,1){3}}
    \put(-8,11){\line(1,0){15}}
  \end{picture}
  \begin{picture}(8,11)
  \end{picture}
)_{ss,\cdot\cdot}=0
\quad,\quad
\sum_{ps}\,p\,(
I^p_\mu
  \begin{picture}(-6,11)
    \put(-6,8){\line(0,1){3}}
    \put(-6,11){\line(1,0){15}}
  \end{picture}
  \begin{picture}(6,11)
  \end{picture}
)_{ss,\cdot\cdot}=0
\label{prop1}\\
\sum_{pp'}(
G^p_\mu
  \begin{picture}(-8,11)
    \put(-8,8){\line(0,1){9}}
    \put(-8,17){\line(1,0){28}}
  \end{picture}
  \begin{picture}(8,11)
  \end{picture}
G^{p'}_{\mu'}
  \begin{picture}(-11,11)
    \put(-11,8){\line(0,1){5}}
    \put(-11,13){\line(1,0){15}}
  \end{picture}
  \begin{picture}(11,11)
  \end{picture}
)_{s\bar{s},\cdot\cdot}=0
\label{prop2}\\
\lim_{D\rightarrow\infty}\sum_{pp'}(
G^p_\mu
  \begin{picture}(-8,11)
    \put(-8,8){\line(0,1){6}}
    \put(-8,14){\line(1,0){25}}
  \end{picture}
  \begin{picture}(8,11)
  \end{picture}
G^{p'}_{\mu'}
  \begin{picture}(-11,11)
    \put(-11,8){\line(0,1){3}}
    \put(-11,11){\line(-1,0){20}}
  \end{picture}
  \begin{picture}(11,11)
  \end{picture}
)_{s\bar{s},\cdot\cdot}=0
\label{prop3}
\end{eqnarray}
The time arguments of the interaction picture are
not written explicitly. The proof can be found in the appendix.

For the reduced density matrix $p(t)$ and the current
$\langle I_L \rangle (t)$, we need the kernels
$\tilde{\Sigma}(z)$ and $\tilde{\Sigma}_I(z)$, see
Eqs.~(\ref{p(z)-solution}) and (\ref{i(z)-solution}).
Due to particle number conservation of the total
Hamiltonian, the reduced density matrix $p(t)$ stays
diagonal if the initial density matrix $p_0$ is
diagonal. We therefore need only the diagonal
matrix elements $\tilde{\Sigma}(z)_{ss,s's'}$ and
$\tilde{\Sigma}_I(z)_{ss,s's'}$ (the nondiagonal
elements can also be calculated but do not contribute
to the current due to $Tr_0$ in Eq.~(\ref{i(z)-solution})).

The kernels are defined in (\ref{sigma}) and 
(\ref{sigma-cur}). Using property (\ref{prop1}) we
find
\begin{equation}
\fbox{\parbox{4.5cm}{
\begin{center}
$\begin{array}{rcl}
\tilde{\Sigma}(z)_{ss,ss} &=& 
- \tilde{\Sigma}(z)_{\bar{s}\bar{s},ss}\,\,, \\
\tilde{\Sigma}_I(z)_{ss,ss} &=& 
\tilde{\Sigma}_I(z)_{\bar{s}\bar{s},ss}\,\,.
\end{array}$
\end{center}}}
\label{exact-relation}
\end{equation}

As a consequence, we only have to calculate the matrix 
element $(11,00)$ of the kernels (the matrix element
$(00,11)$ is analog and we quote the result
at the end).

We call an intermediate double-propagator state
$ss$ "even", and a state $s\bar{s}$ "odd". The vertices
$G$ and $I$ change the parity. We want the matrix
element $(11,00)$, i.e. we start with an even
state $11$ from the left. Thus, the state after
the first vertex of the kernels in
Eqs.~(\ref{sigma}) and (\ref{sigma-cur}) is odd, 
and we can apply properties (\ref{prop2}) and
(\ref{prop3}) for the pair of the second and third 
vertex. We see that the only possibility is that
these two vertices are connected by a pair 
contraction. After these two vertices we are
again in an odd state and can proceed in the
same way. Finally we find that the only
nonvanishing diagrams for $\tilde{\Sigma}(z)$
are
\begin{equation}
A  
\begin{picture}(-2,11)
  \put(-2,8){\line(0,1){8}}
  \put(-2,16){\line(1,0){85}}
  \put(83,8){\line(0,1){8}}
\end{picture}
\begin{picture}(-2,11)
\end{picture}
\,\,\,GG
\begin{picture}(-11,11)
  \put(-11,8){\line(0,1){3}}
  \put(-11,11){\line(1,0){8}}
  \put(-3,8){\line(0,1){3}}
\end{picture}
\begin{picture}(11,11)
\end{picture}
\,\,\,GG
\begin{picture}(-11,11)
  \put(-11,8){\line(0,1){3}}
  \put(-11,11){\line(1,0){8}}
  \put(-3,8){\line(0,1){3}}
\end{picture}
\begin{picture}(11,11)
\end{picture}
\,\dots\,
\,\,\,GG
\begin{picture}(-11,11)
  \put(-11,8){\line(0,1){3}}
  \put(-11,11){\line(1,0){8}}
  \put(-3,8){\line(0,1){3}}
\end{picture}
\begin{picture}(11,11)
\end{picture}
\,B\,\,,\label{dia-left}
\end{equation}
and analog for $\tilde{\Sigma}_I(z)$ by
replacing $A\rightarrow I$.

We denote the sum over all sequences of 
$GG
\begin{picture}(-11,11)
  \put(-11,8){\line(0,1){3}}
  \put(-11,11){\line(1,0){8}}
  \put(-3,8){\line(0,1){3}}
\end{picture}
\begin{picture}(11,11)
\end{picture}$
-blocks in (\ref{dia-left}) by $\Pi(t)$, where $t$
is the time difference between the boundary vertices
$A$ and $B$. $\Pi(t)$ can be calculated analogously
to section \ref{General approach} where we 
resummed sequences of $\Sigma$-blocks with the
result (\ref{p(z)-solution}) for the reduced
density matrix in Laplace space. Thus we obtain
\begin{equation}
\tilde{\Pi}(z)={i\over z-L_0-i\tilde{\sigma}(z)}\,\,,
\end{equation}
with
\begin{equation}
\sigma(t) = GG
\begin{picture}(-11,11)
  \put(-11,8){\line(0,1){3}}
  \put(-11,11){\line(1,0){8}}
  \put(-3,8){\line(0,1){3}}
\end{picture}
\begin{picture}(11,11)
\end{picture}
= \gamma^{pp'}_{\mu\mu'}(t)G^p_\mu e^{-iL_0t}
G^{p'}_{\mu'}\,\,.
\end{equation}
The kernels follow from
\begin{eqnarray}
\Sigma(t) &=&
\gamma^{pp'}_{\mu\mu'}(t)G^p_\mu \Pi(t)
G^{p'}_{\mu'}\,\,,\label{sigma-zw}\\
\Sigma_I(t) &=&
\gamma^{pp'}_{\mu\mu'}(t)I^p_\mu \Pi(t)
G^{p'}_{\mu'}\,\,.\label{sigma-cur-zw}
\end{eqnarray}
Using all definitions together with (\ref{gamma-delta})
and the fact that $L_0$ is a diagonal matrix, we find
in the limit $D\rightarrow\infty$
\begin{eqnarray}
\sigma(t)_{10,10} &=& -\gamma^-(-t)-
\gamma^+(-t)=-\Gamma\,\delta(t)\,\,,
\nonumber\\
\tilde{\sigma}(z)_{10,10} &=& -\Gamma/2\,\,,
\nonumber\\
\tilde{\Pi}(z)_{10,10}&=&{i\over 
z-(L_0)_{10,10}-i\tilde{\sigma}(z)_{10,10}}=
{i\over z-\epsilon+i\Gamma/2}\,\,,
\nonumber\\
\Pi(t)_{10,10}&=& e^{-i\epsilon t} e^{-\Gamma t/2}\,\,,
\end{eqnarray}
with $\gamma^\eta=\sum_r\gamma^\eta_r$ and 
$\Gamma=\sum_r\Gamma_r$. Furthermore, using 
the symmetry relations (\ref{LG-symmetry}) we get
$\Pi(t)_{01,01}=\Pi(t)_{10,10}^*$.
Using these results together with 
$\gamma^\eta_r(t)^*=\gamma^\eta_r(-t)$, we can
evaluate (\ref{sigma-zw}) and
(\ref{sigma-cur-zw}), and find
\begin{eqnarray}
\Sigma(t)_{11,00} &=& 
\gamma^+(t)\Pi(t)_{10,10}\,+\,\mbox{h.c.}
=\gamma^+(t)e^{-i\epsilon t}e^{-\Gamma t/2}
\,+\,\mbox{h.c.}\,\,\\
\Sigma_I(t)_{11,00} &=& 
(e/2)\gamma^+_L(t)\Pi(t)_{10,10}\,+\,\mbox{h.c.}
=(e/2)\gamma_L^+(t)e^{-i\epsilon t}e^{-\Gamma t/2}
\,+\,\mbox{h.c.}
\end{eqnarray}
Using (\ref{gamma-zw}) and (\ref{gamma-eta-def}) we
finally get in Laplace space
\begin{equation}
\fbox{\parbox{10cm}{
\begin{center}
$\begin{array}{rcl}
\tilde{\Sigma}(z)_{11,00}&=&
{i\over 2\pi}\int dE\sum_r\Gamma_r f_r(E)
\left\{{1\over E-\epsilon+z+i\Gamma/2}-
{1\over E-\epsilon-z-i\Gamma/2}\right\}\\
\tilde{\Sigma}_I(z)_{11,00}&=&
{ie\over 4\pi}\int dE\Gamma_L f_L(E)
\left\{{1\over E-\epsilon+z+i\Gamma/2}-
{1\over E-\epsilon-z-i\Gamma/2}\right\}
\end{array}$
\end{center}}}
\label{exact-solution}
\end{equation}
By an analog calculation one obtains the matrix element
$(00,11)$ by replacing $f_r\rightarrow 1-f_r$ 
and changing the sign for the current kernel.

\section{Renormalization group}
\label{Renormalization group}

In this section we will develop a renormalization group
technique to calculate the kernels $\tilde{\Sigma}(z)$
and $\tilde{\Sigma}_I(z)$ in a systematic way beyound
perturbation theory. The two kernels are defined in
(\ref{sigma}) and (\ref{sigma-cur}). Except for the
first boundary vertex, they are formally the same.
Therefore, w.l.o.g. we discuss in the following the
kernel $\tilde{\Sigma}(z)$. Furthermore, as pointed
out after Eq.~(\ref{sigma}), we can set $t_0=0$.

\begin{figure}
  \centerline{\psfig{figure=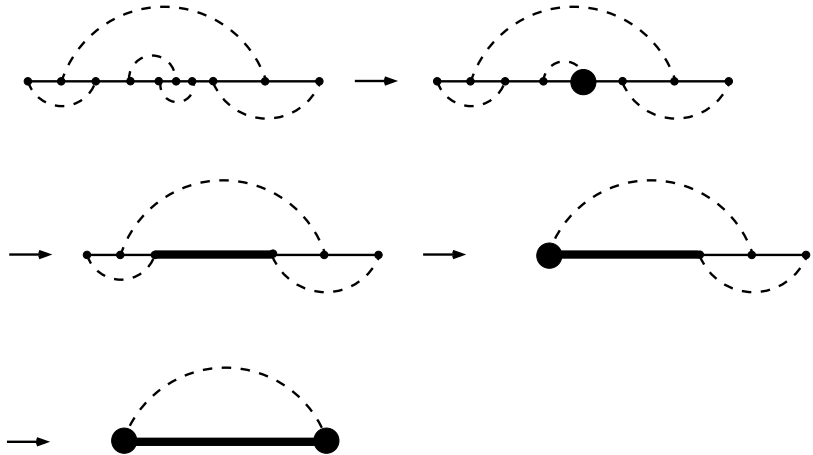,height=6cm}}
  \caption{Illustration of the RG-method. Successively,
the shortest contraction line is integrated out. The thick
lines and dots indicate renormalized propagators and vertices.}
\label{fig-RG}
\end{figure}
\begin{figure}
  \centerline{\psfig{figure=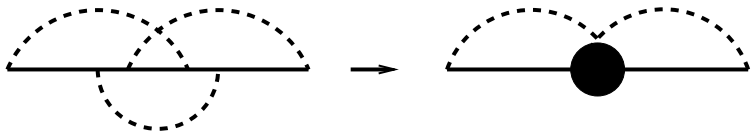,width=10cm}}
  \caption{Generation of a double-vertex.}
\label{fig-double}
\end{figure}

In section \ref{Diagrammatic language} we have found
an effective theory in terms of the dot degrees of
freedom. The reservoirs enter via pair contractions 
which couple the tunneling vertices. The aim is 
to integrate out all contractions in such a
way that they can be interpreted as renormalization
of $L_0$, $G$, $A$ and $B$. The 
procedure is shown schematically in Fig.~\ref{fig-RG}.
We start with the integration over the shortest
contraction. This gives rise to a renormalization
of $G$ in Fig.~\ref{fig-RG}. The integration over
the next shortest contraction renormalizes the propagator
$e^{-iL_0(t_1-t_2)}$ of the dot. Proceeding in the
same way, we find a renormalization of $A$ and $B$ in
the next two steps. Finally we are left with a
diagram which we can calculate easily by perturbation
theory but with renormalized quantities. 
It can also happen that two or more vertices fall into 
one contraction which is integrated out, see
e.g. Fig.~\ref{fig-double}. In this case we generate
double-, triple-, and higher order vertices. In a
perturbative renormalization group treatment one
cuts this infinite hierarchy at a certain level.
Here, we only consider propagator and 
single-vertex renormalization.

As described, we first want to integrate over the short 
contraction lines, i.e. over short time scales 
of the function $\gamma_\mu(t)$. Short time scales
correspond to large energy scales. This leads us
to the usual way renormalization group is
formulated. One introduces
a cut-off function $F(E/D)$ in the integrand
of (\ref{gamma-exact}), where $D$ is a high-energy
cutoff, $F(0)=1$ and $F(x)$ decays monotonically
to zero sufficiently fast for 
$|x|\rightarrow\pm\infty$. As a consequence, the 
pair contraction $\gamma_{\mu,D}(t)$ depends on the 
cutoff $D$. In each renormalization group step
one reduces the cutoff by an infinitesimal
amount $D\rightarrow D-\delta D$, i.e. one
tries to integrate out a small energy shell.
The mathematical challenge is to interpret
the result of this integration as a renormalization
of system parameters, like $L_0$, $G$, $A$ and
$B$. 

\begin{figure}
  \centerline{\psfig{figure=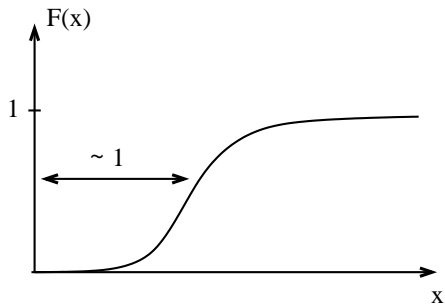,height=4cm}}
  \caption{The cutoff-function.}
\label{fig-cutoff}
\end{figure}

We prefer to develop the renormalization group
procedure in real-time space. This turns out to be easier 
and more systematic. We introduce a cut-off function
$F(t/t_c)$ which cuts off small time scales, 
i.e. $F(x)$, with $x>0$, is a monotonically
increasing function with $F(0)=0$ and 
$\lim_{x\rightarrow \infty}F(x)=1$, 
see Fig.~\ref{fig-cutoff}. $t_c$ is a cutoff
parameter for small time scales. We define a cutoff
dependent reservoir contraction by
\begin{equation}
\gamma_{\mu,t_c}(t) = \gamma_\mu(t)\,F(t/t_c)\,\,.
\label{gamma-cut}
\end{equation}
If we choose a sharp cutoff function $F(x)=\Theta(x-1)$,
$\gamma_{\mu t_c}(t)$ includes only those time scales 
which are precisely larger than $t_c$. We note that the
high-energy cutoff $D$ introduced in 
Eq.~(\ref{gamma-zw}) is independent of $t_c$ and
is not used as a renormalization group flow
parameter. Within our real-time formulation, it 
corresponds to the physical band width of the
reservoirs.

Motivated by the 
picture presented at the beginning of this 
section, we hope that all time scales being smaller
than $t_c$, i.e. those being not present in
$\gamma_{\mu t_c}(t)$ can be accounted for by renormalized 
quantities. To formulate this precisely let us
write the kernel $\tilde{\Sigma}(z)$ as
a functional of $L_0$, $G$, $A$, $B$ and 
$\gamma$
\begin{equation}
\tilde{\Sigma}(z)={\cal{F}}
(L_0,G^p_\mu,A^p_\mu,B^p_\mu,\gamma_\mu(t))\,\,,
\end{equation}
where $p$, $\mu$,and $t$ run over all possible
values within the functional.
After replacing $\gamma_\mu$ by the cutoff-dependent
function $\gamma_{\mu t_c}$ inside this functional,
we try to find a cutoff dependence of all
other quantities in such a way that the kernel
stays invariant with the \underline{{\it same}} 
functional ${\cal{F}}$
\begin{equation}
\tilde{\Sigma}(z) = \tilde{\Sigma}_{t_c}(z)
\,+\,{\cal{F}}(L_{0t_c},G^p_{\mu t_c},A^p_{\mu t_c},
B^p_{\mu t_c},\gamma_{\mu t_c}(t))\,\,.
\label{invariance}
\end{equation}
This equation is only exact
if we neglect higher order vertex corrections, as
already pointed out above. Otherwise we have to
include all higher-order vertex terms as arguments
in the functional ${\cal{F}}$ as well.
Via $\gamma_{\mu t_c}$, the second term on the r.h.s 
includes only time scales which are larger
than $t_c$. All other time scales are accounted for
by the cutoff-dependence of $L_0$, $G$, $A$, and
$B$. The first term on the r.h.s contains the 
contributions where {\it all} time scales are smaller 
than $t_c$. This part is not included in the second
term where at least one contraction line
$\gamma_{\mu t_c}$ occurs. It is important
to notice that $\tilde{\Sigma}_{t_c}(z)$ should
not be viewed as being trivially defined by 
Eq.~(\ref{invariance}). If we choose a sharp cutoff
function $F(x)=\Theta(x-1)$,  
$\tilde{\Sigma}_{t_c}(z)$ can
be contructively defined as containing only
those diagrams where {\it all} contraction lines
have a length smaller than $t_c$. For 
$t_c\rightarrow\infty$ the second term on the
r.h.s. of (\ref{invariance}) vanishes, provided
the function $\gamma_\mu(t)$ decays sufficiently
fast for $t\rightarrow\infty$. Thus, we obtain
the final solution from
\begin{equation}
\tilde{\Sigma}(z) = \lim_{t_c\rightarrow\infty}
\tilde{\Sigma}_{t_c}(z)\,\,.
\label{sigma-asymp}
\end{equation}
The aim is to find differential equations which
describe the cutoff-dependence of 
$\tilde{\Sigma}_{t_c}(z)$, $L_{0t_c}$, $G_{t_c}$, 
$A_{t_c}$ and $B_{t_c}$. This will be the subject 
of the rest of this section.

Let us increase the cutoff by an infinitesimal
amount $t_c\rightarrow t_c+dt_c$. Due to the
invariance, Eq.~(\ref{invariance}) should not
change
\begin{equation}
0\,=\,d\tilde{\Sigma}(z)
\,+\,{\cal{F}}(L_0+dL_0,G+dG,A+dA,B+dB,
\gamma+d\gamma)\,\,,
\label{change}
\end{equation}
where we have omitted the subindex $t_c$ and the
indizes $p$ and $\mu$. The change of $\gamma$ is
known from the definition (\ref{gamma-cut})
\begin{equation}
d\gamma_{\mu t_c}(t) = - \gamma_\mu(t)
F^\prime(t/t_c)t/t_c^2\,dt_c\,\,,
\end{equation}
or, for a sharp cutoff function $F(x)=\Theta(x-1)$
\begin{equation}
d\gamma_{\mu t_c}(t)= -\gamma_\mu(t_c)\delta(t-t_c)
\,dt_c\,\,.
\label{change-sharp}
\end{equation}
The increment $d\gamma$ in (\ref{change}) leads 
to all diagrams where one contraction line is 
replaced by $d\gamma$. We indicate this by a
cross and define
\begin{equation}
  G_1\dots G_2 
  \begin{picture}(-35,11) 
    \put(-35,8){\line(0,1){3}} 
    \put(-35,11){\line(1,0){27}} 
    \put(-8,8){\line(0,1){3}}
    \put(-25,8.5){$\times$}
  \end{picture}
  \begin{picture}(30,11) 
  \end{picture}\,\,\,\,
= \,-\,\,
({d\gamma\over dt_c})^{pp'}_{\mu\mu'}(t_1-t_2)\,
\,dt_c\,
G^p_\mu(t_1)\dots G^{p'}_{\mu'}(t_2)\,\,,
\label{cross}
\end{equation}
We call this a "cross contraction". 
For a sharp cutoff function, we get
\begin{equation}
  G_1\dots G_2 
  \begin{picture}(-35,11) 
    \put(-35,8){\line(0,1){3}} 
    \put(-35,11){\line(1,0){27}} 
    \put(-8,8){\line(0,1){3}}
    \put(-25,8.5){$\times$}
  \end{picture}
  \begin{picture}(30,11) 
  \end{picture}\,\,\,\,
= \,\gamma^{pp'}_{\mu\mu'}(t_c)\,
\delta(t_1-t_2-t_c)\,dt_c\,
G^p_\mu(t_1)\dots G^{p'}_{\mu'}(t_2)\,\,.
\end{equation}
Corresponding definitions hold for 
$G$ replaced by $A$ or $B$. By convention
we have included a minus sign into the
definition since $-d\gamma$ accounts for
the time scales of the contraction 
between $t_c$ and $t_c+dt_c$, compare
Eq.~(\ref{change-sharp}).
We have to identify all terms created
by $d\gamma$ with a contribution arising either 
from $d\tilde{\Sigma}$, $dL_0$, $dG$, $dA$ or 
$dB$ in Eq.~(\ref{change}) in order to
fulfil invariance. Which one
has to be taken depends on the number
of vertices between $G_1$ and $G_2$, and
wether one or both of the vertices 
are boundary vertices. 

Let us start with the simplest case of 
two successive vertices which are not at
the boundary
\begin{equation}
G_1G_2G_3 
  \begin{picture}(-20,11) 
    \put(-20,8){\line(0,1){3}} 
    \put(-20,11){\line(1,0){12}} 
    \put(-8,8){\line(0,1){3}}
    \put(-17.5,8.5){$\times$}
  \end{picture}
  \begin{picture}(20,11) 
  \end{picture}
G_4\,\,.
\end{equation}
Here, the two vertices $G_1$ and $G_4$ are 
contracted to any other vertices. Motivated
by Fig.~\ref{fig-RG}, it is tempting to
take this diagram together with $G_1G_4$
and interprete the cross contraction
as a renormalization of $L_0$ in the
following sense
\begin{eqnarray}
G_1G_4\quad +
\int\limits_{1>2>3>4}dt_2dt_3\,
G_1G_2G_3 
  \begin{picture}(-20,11) 
    \put(-20,8){\line(0,1){3}} 
    \put(-20,11){\line(1,0){12}} 
    \put(-8,8){\line(0,1){3}}
    \put(-17.5,8.5){$\times$}
  \end{picture}
  \begin{picture}(20,11) 
  \end{picture}
G_4\,\nonumber\\
\quad
\begin{array}[b]{c}
\vspace{-1.5mm}?\\=\end{array}
\quad e^{iL_0t_1}G^{p_1}_{\mu_1}e^{-i(L_0+dL_0)(t_1-t_4)}
G^{p_4}_{\mu_4}e^{-iL_0t_4}\,\,.
\end{eqnarray}
Expanding the exponential to linear order in
$dL_0$, we see that this requires the identity
\begin{equation}
\int\limits_{1>2>3>4}dt_2dt_3\,
G_1G_2G_3 
  \begin{picture}(-20,11) 
    \put(-20,8){\line(0,1){3}} 
    \put(-20,11){\line(1,0){12}} 
    \put(-8,8){\line(0,1){3}}
    \put(-17.5,8.5){$\times$}
  \end{picture}
  \begin{picture}(20,11) 
  \end{picture}
G_4\quad 
\begin{array}[b]{c}
\vspace{-1.5mm}?\\=\end{array}
\quad -i\,\int\limits_{t_1>t>t_4}dt\,
G_1(dL_0)(t)G_4\,\,.
\label{question}
\end{equation}
To obtain such a relation we see immediately
a problem. The integrand
of the l.h.s contains a two-time object 
$G_2G_3 
  \begin{picture}(-20,11) 
    \put(-20,8){\line(0,1){3}} 
    \put(-20,11){\line(1,0){12}} 
    \put(-8,8){\line(0,1){3}}
    \put(-17.5,8.5){$\times$}
  \end{picture}
  \begin{picture}(20,11) 
  \end{picture}$, 
whereas the integrand of the r.h.s
contains the single-time object
$(dL_0)(t)$. We have to decide
to which time-variable we identify $t$.
Let us make an arbitrary choice: $t\equiv t_3$.
Then it is obvious to try the ansatz
\begin{equation}
-i\,(dL_0)(t_3)\,=\,\int\limits_{2>3} dt_2
G_2G_3 
  \begin{picture}(-20,11) 
    \put(-20,8){\line(0,1){3}} 
    \put(-20,11){\line(1,0){12}} 
    \put(-8,8){\line(0,1){3}}
    \put(-17.5,8.5){$\times$}
  \end{picture}
  \begin{picture}(20,11) 
  \end{picture}
\,\,.
\label{L-ansatz}
\end{equation}
This is well-defined since it does not involve
the time variables $t_1$ and $t_4$. However,
let us try to insert this ansatz into the
r.h.s of (\ref{question}). We get two terms
\begin{eqnarray}
-i\,\int\limits_{1>3>4}dt_3\,G_1(dL_0)(t_3)G_4&=&
\int\limits_{1>3>4\atop 2>3} dt_2dt_3
\,\,G_1G_2G_3 
  \begin{picture}(-20,11) 
    \put(-20,8){\line(0,1){3}} 
    \put(-20,11){\line(1,0){12}} 
    \put(-8,8){\line(0,1){3}}
    \put(-17.5,8.5){$\times$}
  \end{picture}
  \begin{picture}(20,11) 
  \end{picture}G_4
\nonumber\\
&&\hspace{-3cm}=
\int\limits_{1>2>3>4} dt_2dt_3
\,\,G_1G_2G_3 
  \begin{picture}(-20,11) 
    \put(-20,8){\line(0,1){3}} 
    \put(-20,11){\line(1,0){12}} 
    \put(-8,8){\line(0,1){3}}
    \put(-17.5,8.5){$\times$}
  \end{picture}
  \begin{picture}(20,11) 
  \end{picture}G_4
\quad +
\int\limits_{2>1>3>4} dt_2dt_3
\,\,G_1G_2G_3 
  \begin{picture}(-20,11) 
    \put(-20,8){\line(0,1){3}} 
    \put(-20,11){\line(1,0){12}} 
    \put(-8,8){\line(0,1){3}}
    \put(-17.5,8.5){$\times$}
  \end{picture}
  \begin{picture}(20,11) 
  \end{picture}G_4\,\,.
\end{eqnarray}
The first term agrees with the l.h.s. of 
(\ref{question}), whereas the second one
is a correction term which has to be considered
separately. Let us try to interpret it as
a term arising from $dG$. We define
\begin{equation}
(dG^{p_1}_{\mu_1})^{(1)}(t_1)\,\, =
\,-\int\limits_{2>1>3}
dt_2dt_3\,G_1
G_2G_3 
  \begin{picture}(-20,11) 
    \put(-20,8){\line(0,1){3}} 
    \put(-20,11){\line(1,0){12}} 
    \put(-8,8){\line(0,1){3}}
    \put(-17.5,8.5){$\times$}
  \end{picture}
  \begin{picture}(20,11) 
  \end{picture}\,\,,
\label{G-ansatz}
\end{equation}
where the subindex $(1)$ indicates that
this is the first contribution to $dG$
(another one will follow below). Using
this definition we finally get
\begin{eqnarray}
\int\limits_{1>2>3>4}dt_2dt_3\,G_1
G_2G_3 
  \begin{picture}(-20,11) 
    \put(-20,8){\line(0,1){3}} 
    \put(-20,11){\line(1,0){12}} 
    \put(-8,8){\line(0,1){3}}
    \put(-17.5,8.5){$\times$}
  \end{picture}
  \begin{picture}(20,11) 
  \end{picture}G_4
\,&=&\,-i\,\int\limits_{1>3>4}dt_3\,G_1
(dL_0)(t_3)G_4\,+\,(dG^{p_1}_{\mu_1})^{(1)}(t_1)G_4
\nonumber\\
&&+\int\limits_{2>1>4>3}dt_2dt_3
\,G_1G_2G_3 
  \begin{picture}(-20,11) 
    \put(-20,8){\line(0,1){3}} 
    \put(-20,11){\line(1,0){12}} 
    \put(-8,8){\line(0,1){3}}
    \put(-17.5,8.5){$\times$}
  \end{picture}
  \begin{picture}(20,11) 
  \end{picture}G_4\,\,.
\label{final-res}
\end{eqnarray}
In the last term on the r.h.s. the two
time-variables $t_1$ and $t_4$ are
"clustered" together, they both have
to lie within the time interval 
$[t_3,t_2]$. Therefore, we interpret
this term as a double-vertex which 
we neglect. In conclusion, we have achieved 
our final goal: we have interpreted a term
arising from $d\gamma$ by a
renormalization of $L_0$ and $G$.

Let us try to understand what we have
done so far and how to find a systematic
procedure for obtaining the complete
RG equations without going again into
the details of the foregoing exercise.
We have defined the renormalization of
$L_0$ in (\ref{L-ansatz}), which is
natural since this expression contains
only operators of the dot. We write
\begin{equation}
-i\,(dL_0)(t_2)\,=\,
G_1G_2
  \begin{picture}(-20,11) 
    \put(-20,8){\line(0,1){3}} 
    \put(-20,11){\line(1,0){12}} 
    \put(-8,8){\line(0,1){3}}
    \put(-17.5,8.5){$\times$}
    \put(-5,-10){$\uparrow$}
  \end{picture}
  \begin{picture}(20,11) 
  \end{picture}\,\,,
\label{L-ren}
\end{equation}

\vspace{1mm}
\noindent
where the arrow indicates the time
variable which is not integrated out.
It is {\it this} time variable which is
used for the definition of the 
interaction picture of $dL_0$ and,
most importantly, which is used for
the definition of time-ordering, see
the r.h.s of Eq.~(\ref{question}).
The time-variable $t_1$ no longer
appears explicitly because it is
an internal integration variable
within the definition of $dL_0$.
This has the consequence, that time
variables of other  
vertices do not care about the value
of $t_1$ regarding time ordering. E.g.,
if we multiply $(dL_0)(t_3)$ with
a vertex $G_2$ from the left, 
time-ordering only requires 
$t_2>t_3$ but the ordering of $t_2$
with respect to the internal integration
variable $t_1$ is not prescribed. Thus,
due to $dL_0$, the following term will 
occur on the r.h.s. of (\ref{change}) 
\begin{equation}
  G_2G_1G_3 
  \begin{picture}(-20,11) 
    \put(-20,8){\line(0,1){3}} 
    \put(-20,11){\line(1,0){12}} 
    \put(-8,8){\line(0,1){3}}
    \put(-17.5,8.5){$\times$}
    \put(-30,-10){$\uparrow$}
    \put(-5,-10){$\uparrow$}
  \end{picture}
  \begin{picture}(20,11) 
  \end{picture}\,\,,
\label{correction-term}
\end{equation}
where $t_1>t_2>t_3$. However, this
term is not present in the original
series because the ordering of operators
does not agree with the ordering of
time variables. Consequently, we have
to subtract this term. This is the
basic reason for the occurence of
the second term on the r.h.s. of
(\ref{final-res}). 

We interpret minus the correction term 
(\ref{correction-term}) as renormalization
of $G$. Again we have to specify the
time variable for the interaction picture
and for time-ordering. We
choose $t_2$ in analogy to
Eq.~(\ref{G-ansatz}).
Together with the "real" vertex
renormalization, arising from a free
vertex inside a cross contraction, we
get the complete renormalization of
$G$
\begin{equation}
(dG^{p_2}_{\mu_2})(t_2)\,=\,
G_1G_2G_3
  \begin{picture}(-33,11) 
    \put(-33,8){\line(0,1){3}} 
    \put(-33,11){\line(1,0){25}} 
    \put(-8,8){\line(0,1){3}}
    \put(-24,8.5){$\times$}
    \put(-17,-10){$\uparrow$}
  \end{picture}
  \begin{picture}(33,11) 
  \end{picture}
\,-\,  G_2G_1G_3 
  \begin{picture}(-20,11) 
    \put(-20,8){\line(0,1){3}} 
    \put(-20,11){\line(1,0){12}} 
    \put(-8,8){\line(0,1){3}}
    \put(-17.5,8.5){$\times$}
    \put(-30,-10){$\uparrow$}
  \end{picture}
  \begin{picture}(20,11) 
  \end{picture}\,\,.
\label{G-ren}
\end{equation}
We can now proceed to look at other correction 
terms which can occur by multiplying 
(\ref{L-ren}) with two vertices from the left
or (\ref{G-ren}) with one vertex from the left
or right
\begin{equation}
G_2G_3G_1G_4
  \begin{picture}(-20,11) 
    \put(-20,8){\line(0,1){3}} 
    \put(-20,11){\line(1,0){12}} 
    \put(-8,8){\line(0,1){3}}
    \put(-17.5,8.5){$\times$}
    \put(-5,-10){$\uparrow$}
    \put(-30,-10){$\uparrow$}
    \put(-42,-10){$\uparrow$}
  \end{picture}
  \begin{picture}(20,11) 
  \end{picture}\,+\,
G_2G_1G_3G_4
  \begin{picture}(-33,11) 
    \put(-33,8){\line(0,1){3}} 
    \put(-33,11){\line(1,0){25}} 
    \put(-8,8){\line(0,1){3}}
    \put(-24,8.5){$\times$}
    \put(-17,-10){$\uparrow$}
    \put(-42,-10){$\uparrow$}
  \end{picture}
  \begin{picture}(33,11) 
  \end{picture}
\,-\,  G_2G_3G_1G_4 
  \begin{picture}(-20,11) 
    \put(-20,8){\line(0,1){3}} 
    \put(-20,11){\line(1,0){12}} 
    \put(-8,8){\line(0,1){3}}
    \put(-17.5,8.5){$\times$}
    \put(-30,-10){$\uparrow$}
    \put(-42,-10){$\uparrow$}
  \end{picture}
  \begin{picture}(20,11) 
  \end{picture}\,+\,
G_1G_2G_4
  \begin{picture}(-33,11) 
    \put(-33,8){\line(0,1){3}} 
    \put(-33,11){\line(1,0){25}} 
    \put(-8,8){\line(0,1){3}}
    \put(-24,8.5){$\times$}
    \put(-17,-10){$\uparrow$}
  \end{picture}
  \begin{picture}(33,11) 
  \end{picture}
G_3
  \begin{picture}(-10,11) 
    \put(-5,-10){$\uparrow$}
  \end{picture}
  \begin{picture}(10,11) 
  \end{picture}
\,-\,  G_2G_1G_4
  \begin{picture}(-20,11) 
    \put(-20,8){\line(0,1){3}} 
    \put(-20,11){\line(1,0){12}} 
    \put(-8,8){\line(0,1){3}}
    \put(-17.5,8.5){$\times$}
    \put(-30,-10){$\uparrow$}
  \end{picture}
  \begin{picture}(20,11) 
  \end{picture}
G_3
  \begin{picture}(-10,11) 
    \put(-5,-10){$\uparrow$}
  \end{picture}
  \begin{picture}(10,11) 
  \end{picture}\,\,,
\end{equation}
with $t_1>t_2>t_3>t_4$. We have indicated
the time-ordering variables of all sub-clusters
which lead to these expressions. We see that 
the first and third term cancel each other. The
other three correction terms have to
be subtracted again. Minus the fifth term
corresponds to the third term on the
r.h.s. of Eq.~(\ref{final-res}). The
second and fourth term are correction
terms arising from the "real" vertex 
renormalization. However, all these
correction terms correspond to 
double-vertices which we neglect 
consistently. Another double-vertex
term arises from two free vertices 
within a cross contraction. Proceeding
further in the same way, triple- and
higher order vertex terms will occur
which again are not considered here.
We see that the procedure is very 
systematic and straightforward.

In the same way we can proceed for
boundary vertices. Since we want to
calculate the Laplace transform
$\tilde{\Sigma}(z)$ of the kernel
each diagram of the kernel (\ref{sigma}) 
gets an additonal factor $e^{izt}$.
Consequently we define the interaction
picture for the boundary vertices 
slightly different from (\ref{AB-intpic}).
We simply include the factor from the
Laplace transformation
\begin{equation}
A^p_\mu(t) = e^{izt}A^p_\mu e^{-iL_0t} 
\quad,\quad
B^p_\mu(t) = e^{iL_0t} B^p_\mu e^{-izt}\,\,.
\label{AB-intpic-z}
\end{equation}
Analog to the above procedure we can write
down immediately all terms with
cross contractions containing boundary
operators. They can be interpreted as
renormalization of $\tilde{\Sigma}$,
$A$ and $B$
\begin{eqnarray}
\hspace{1cm} d\tilde{\Sigma}(z)\,&=&\,
A_1B_2
  \begin{picture}(-20,11) 
    \put(-20,8){\line(0,1){3}} 
    \put(-20,11){\line(1,0){12}} 
    \put(-8,8){\line(0,1){3}}
    \put(-17.5,8.5){$\times$}
  \end{picture}
  \begin{picture}(20,11) 
  \end{picture}
|_{t_2=0}
\,\,,
\label{sigma-ren}\\\nonumber\\
(dA^{p_2}_{\mu_2})(t_2)\,&=&\,
A_1G_2G_3
  \begin{picture}(-33,11) 
    \put(-33,8){\line(0,1){3}} 
    \put(-33,11){\line(1,0){25}} 
    \put(-8,8){\line(0,1){3}}
    \put(-24,8.5){$\times$}
    \put(-17,-10){$\uparrow$}
  \end{picture}
  \begin{picture}(33,11) 
  \end{picture}
\,-\,  A_2G_1G_3 
  \begin{picture}(-20,11) 
    \put(-20,8){\line(0,1){3}} 
    \put(-20,11){\line(1,0){12}} 
    \put(-8,8){\line(0,1){3}}
    \put(-17.5,8.5){$\times$}
    \put(-30,-10){$\uparrow$}
  \end{picture}
  \begin{picture}(20,11) 
  \end{picture}\,\,,
\label{A-ren}\\ \nonumber\\
(dB^{p_2}_{\mu_2})(t_2)\,&=&\,
G_1G_2B_3
  \begin{picture}(-33,11) 
    \put(-33,8){\line(0,1){3}} 
    \put(-33,11){\line(1,0){25}} 
    \put(-8,8){\line(0,1){3}}
    \put(-24,8.5){$\times$}
    \put(-17,-10){$\uparrow$}
  \end{picture}
  \begin{picture}(33,11) 
  \end{picture}\,\,.
\label{B-ren}
\end{eqnarray}
Again, we integrate implicitly over
all time variables which are not indicated
by an arrow, with $t_1>t_2>t_3$. Terms like
$  A_1G_2 
  \begin{picture}(-20,11) 
    \put(-20,8){\line(0,1){3}} 
    \put(-20,11){\line(1,0){12}} 
    \put(-8,8){\line(0,1){3}}
    \put(-17.5,8.5){$\times$}
  \end{picture}
  \begin{picture}(20,11) 
  \end{picture}$ or
$  G_1B_2 
  \begin{picture}(-20,11) 
    \put(-20,8){\line(0,1){3}} 
    \put(-20,11){\line(1,0){12}} 
    \put(-8,8){\line(0,1){3}}
    \put(-17.5,8.5){$\times$}
  \end{picture}
  \begin{picture}(20,11) 
  \end{picture}$
do not occur since they do not lead to connected
diagrams. For this reason, there is also no
correction term in the renormalization group
equation for $B$.

\begin{figure}
  \centerline{\psfig{figure=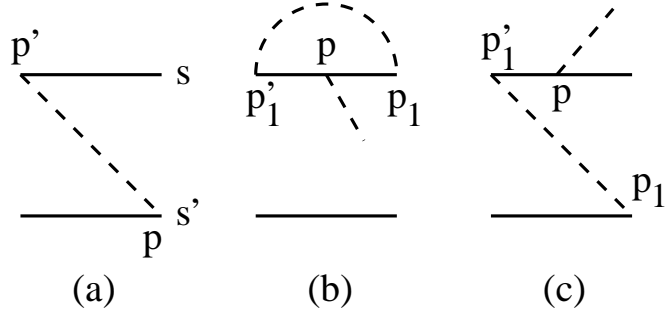,height=4cm}}
  \caption{Diagrams where additional minus occur if the
ends of the indicated contraction are connected. In (a) a
minus signs occurs if the particle number difference 
$N_s-N_{s'}$ is odd. In (b) and (c) a free vertex
is crossed by connecting the end-points
without crossing over $p_0$.}
\label{fig-sign}
\end{figure}
Eqs.~(\ref{L-ren}), (\ref{G-ren}), and 
(\ref{sigma-ren})-(\ref{B-ren}) are the final
RG-equations. They can be translated easily
to get explicit expressions. In order to
get the renormalization of the bare quantities
we have to set $t_2=0$ in all equations.
To account for possible minus signs arising
from commutation of fermionic reservoir field
operators, we have to include two auxiliary
sign functions. They arise because we have to
connect the two end points of the cross
contraction. If the cross contraction couples
the upper with the lower propagator, a minus
sign occurs if the intermediate
double-propagator state at the vertex with
the larger time is odd (i.e. if the difference
of the fermionic particle numbers on the
upper and lower propagator is odd), see
Fig.~\ref{fig-sign}a. We account
this by a sign operator $\hat{\sigma}^{pp'}_\mu$ 
multiplying each cross contraction from
the left. The matrix elements of this 
operator are defined by
\begin{equation}
(\hat{\sigma}^{pp'}_\mu)_{ss',ss'}\,=\,
\left\{
\begin{array}{ccl}
pp'&\mbox{for}&N_s-N_{s'}=\mbox{odd}\\
1&\mbox{for}&N_s-N_{s'}=\mbox{even}
\end{array}
\right.\,\,,
\end{equation}
where $N_s$ is the fermionic particle number 
of state $s$ and $\mu$ is a fermionic
contraction (for bosonic contractions no
additional sign has to be considered).
The second sign function concerns the
case when one free vertex occurs within
a cross contraction.
A minus sign occurs when the cross contraction
crosses over the free vertex when we want
to connect the two end points. Since
$\hat{\sigma}$ is multiplied from the left
by definition, we connect the vertices always
in such a way that we do not cross over
$p_0$. We denote by $p$, $p_1$, and $p_1^\prime$ 
the indices of the free vertex, the vertex
with the larger and the one with the 
smaller time variable
of the cross contraction, respectively.
If $p=p_1^\prime$ we get an additional
minus sign, see 
Fig.~\ref{fig-sign}b and \ref{fig-sign}c.
Therefore we multiply the "real" vertex
correction by an additional sign function
\begin{equation}
\eta^{pp_1^\prime}_{\mu\mu_1^\prime}\,=\,
\left\{
\begin{array}{cl}
-pp_1^\prime&\mbox{for }\mu \mbox{ and }
\mu' \mbox{ fermionic}\\
1&\mbox{otherwise}
\end{array}
\right.\,\,.
\end{equation}
In summary we get the following RG equations
\begin{equation}
\fbox{\parbox{12cm}{
\begin{center}
$\begin{array}{rcl}
   {d\over dt_c}\tilde{\Sigma}(z)&=&
   -\int\limits_0^\infty dt\,
   ({d\gamma\over dt_c})^{pp'}_{\mu\mu'}(t)
   \hat{\sigma}^{pp'}_\mu A^p_\mu(t)B^{p'}_{\mu'}
   \\ \\
   {d\over dt_c}L_0&=&
   -i\int\limits_0^\infty dt\,
   ({d\gamma\over dt_c})^{pp'}_{\mu\mu'}(t)
   \hat{\sigma}^{pp'}_\mu G^p_\mu(t)G^{p'}_{\mu'}\\ \\
   {d\over dt_c}G^p_\mu&=&
   -\int\limits_0^\infty dt\int\limits_{-\infty}^0 dt'\,
   ({d\gamma\over dt_c})^{p_1p^\prime_1}_{\mu_1\mu^\prime_1}(t-t')
   \nonumber\\
   &&\hspace{1.5cm}\left[\eta^{pp^\prime_1}_{\mu\mu_1}
   \hat{\sigma}^{p_1p^\prime_1}_{\mu_1} G^{p_1}_{\mu_1}(t)
   G^p_\mu\, - \,G^p_\mu \hat{\sigma}^{p_1p^\prime_1}_{\mu_1} 
   G^{p_1}_{\mu_1}(t)\,\right]\,
   G^{p_1^\prime}_{\mu_1^\prime}(t')\\ \\
   {d\over dt_c}A^p_\mu&=&
   -\int\limits_0^\infty dt\int\limits_{-\infty}^0 dt'\,
   ({d\gamma\over dt_c})^{p_1p^\prime_1}_{\mu_1\mu^\prime_1}(t-t')
   \nonumber\\
   &&\hspace{1.5cm}\left[\,\eta^{pp^\prime_1}_{\mu\mu_1}
   \hat{\sigma}^{p_1p^\prime_1}_{\mu_1} A^{p_1}_{\mu_1}(t)
   G^p_\mu \,- \,A^p_\mu \hat{\sigma}^{p_1p^\prime_1}_{\mu_1} 
   G^{p_1}_{\mu_1}(t)\,\right]\,
   G^{p_1^\prime}_{\mu_1^\prime}(t')\\ \\
   {d\over dt_c}B^p_\mu&=&
   -\int\limits_0^\infty dt\int\limits_{-\infty}^0 dt'\,
   ({d\gamma\over dt_c})^{p_1p^\prime_1}_{\mu_1\mu^\prime_1}(t-t')
   \,\eta^{pp^\prime_1}_{\mu\mu_1}
   \hat{\sigma}^{p_1p^\prime_1}_{\mu_1} G^{p_1}_{\mu_1}(t)
   G^p_\mu B^{p_1^\prime}_{\mu_1^\prime}(t')
\end{array}$
\end{center}}}
\label{rg-equations}
\end{equation}

\vspace{3mm}
\noindent
Implicitly we sum over all double indices on the r.h.s. of
these equations which do not occur on the l.h.s..
We note that the overall sign on the r.h.s.
differs from the one in \cite{RG-anderson} since we have
included the factors $-i$ into the vertices. Furthermore,
we have generalized the RG equations to an arbitrary
cutoff-function $F(t/tc)$. For
a specific cutoff-function, e.g. $F(x)=\theta(x-1)$, and
taking matrix elements of the RG-equations, all integrals
in (\ref{rg-equations}) can be calculated analytically.
We are left with pure differential equations which can
be solved numerically in a straightforward and very
efficient way. Finally, the asymptotic value
$\lim_{t_c\rightarrow\infty}\tilde{\Sigma}_{t_c}(z)$
gives the solution for the kernel, see 
Eq.~(\ref{sigma-asymp}). Using (\ref{p(z)-solution})
and (\ref{p(t)-solution}), we get the complete
time-evolution of the reduced density matrix of
the dot. The initial condition for the boundary
operators at $t_c=0$ is given by $A^p_\mu=B^p_\mu=G^p_\mu$.
To get the current kernel $\tilde{\Sigma}_I(z)$, we
take the same RG-equations with the initial conditions
$A^p_\mu=I^p_\mu$ and $B^p_\mu=G^p_\mu$.

Numerically, one can not take
the initial flow parameter $t_c^0=0$ and the final one
$t_c^f\rightarrow\infty$. However, one can check
that the final solution is stabel
for sufficiently small $t_c^0$ and large $t_c^f$.
This means that $1/{t_c^0}$ and $1/{t_c^f}$ have to
be much larger resp. smaller than all other energy
scales 
\begin{equation}
{1\over t_c^f}\ll \Gamma_L,\Gamma_R,T,eV,
|\epsilon|\ll D \ll {1\over t_c^0}\,\,,
\label{conditions}
\end{equation}
where $eV=\mu_L-\mu_R$ is the bias voltage, and
$\Gamma_L,\Gamma_R$ are defined in (\ref{Gamma-def}).
In addition, we want to send the band-width cutoff
$D\rightarrow\infty$. This means that we
have to check the stability of the solution
for $t_c^0\rightarrow 0$, $D\rightarrow\infty$,
with $Dt_c^0\rightarrow 0$. For the problem under
consideration this is indeed possible.

There are three essential differences of our final RG 
equations to conventional poor man scaling and 
operator product expansion techniques \cite{hewson,aff-lud}:
\begin{center}
\fbox{\hspace{0.5cm}\parbox{11cm}{\vspace{0.5cm}
{\bf 1.} We have formulated the RG-equations within
a real-time formalism on the Keldysh-contour. 
\\ \\
{\bf 2.} We have not expanded the interaction 
picture of the vertex operators for small times,
i.e. the renormalized propagation $\exp(\pm iL_0 t)$ 
is taken fully into account.
\\ \\
{\bf 3.} For $t_c>t_c^0$ we get $L_0\ne [H_0,\cdot]$,
i.e. we generate non-Hamiltonian dynamics during RG. 
\vspace{0.5cm}
}\hspace{0.5cm}}
\end{center}
\vspace{0.5cm}
\noindent
The first point is necessary for describing
nonequilibrium phenomena. The second one is important
for self-consistency reasons ($L_0$ in the exponent is a
renormalized quantity) and for the stability of
the solution for $t_c^f\rightarrow\infty$. Only
if one expands the exponentials $\exp(\pm i L_0 t)$
in $t$, one has to stop the RG-flow when
$|\lambda t_c|\sim 1$, where $\lambda$ denotes any
eigenvalue of $L_0$. We do not need this expansion
and, consequently, are able to integrate out all
time scales. The third property
is the most important one. It is essential
for a nonequilibrium theory to describe the physics
of dissipation. Thus, the renormalized superoperator
$L_0$ should no longer be expressible by a
commutator with a renormalized Hamiltonian $H_0$, 
i.e. in matrix notation
\begin{equation}
(L_0)_{s_1s_2,s_1's_2'}\,\ne\,
(H_0)_{s_1s_2}\delta_{s_1's_2'}\,-\,
\delta_{s_1s_2}(H_0)_{s_2's_1'}\,\,.
\label{L0-ham1}
\end{equation}
To show this let us first define the renormalized
Hamiltonian in a natural way. It is obvious that
contractions which connect the upper with the
lower propagator of the Keldysh contour, do not lead 
to Hamiltonian 
dynamics. This was already explained at the end of
section \ref{Superoperator notation}. Thus, in
order to define the renormalized $H_0$ let us
consider the same RG equations but allow only
for contractions within the upper or lower
propagator. Formally, this means $p=p'$ 
and $p_1=p_1'$ in (\ref{rg-equations}).
Under these conditions we can write the 
solution as
\begin{equation}
(L_0)_{s_1s_2,s_1's_2'}\,=\,
(H_0)_{s_1s_2}\delta_{s_1's_2'}\,-\,
\delta_{s_1s_2}(H_0^\dagger)_{s_2's_1'}\,\,.
\label{L0-ham2}
\end{equation}
To obtain this, we have used the symmetry 
relations (\ref{LG-symmetry}) together with the property
\begin{equation}
\gamma^{pp'}_{\mu\mu'}(t)^* =
\gamma^{pp'}_{\mu\mu'}(-t)  =
\gamma^{\bar{p}\bar{p}'}_{\bar{\mu}\bar{\mu}'}(t)\,\,.
\end{equation}
The renormalized Hamiltonian $H_0$ is defined by 
considering only the upper propagator, i.e.
setting $p=p'=p_1=p_1'=+$ and replacing
$L_0\rightarrow H_0$ and
$G^+_\mu\rightarrow -ig_\mu$ in the RG-equations.
We note that $H_0$ is non-hermitian since we
generate complex energies during RG. Physically
this describes broadening of levels or finite
life-times. From (\ref{L0-ham2}) we conclude
that the backward propagator involves $H_0^\dagger$,
i.e. the sign of energy-broadening is different
for the forward and backward propagator.
This means, that even if we disregard contractions connecting
the upper with the lower propagator, 
we get $L_0\ne [H_0,\cdot]$. However, this does not
imply the physics of dissipation. It describes 
the physics of a finite broadening of the energy levels
due to the coupling to the environment. This reflects
the Heisenberg-uncertainty relationship.

In contrast,
if we couple the forward and backward propagator
by contractions, the matrix $L_0$ will
completely change its form. It can neither be represented
as (\ref{L0-ham1}) nor as (\ref{L0-ham2}). The coupling
of the propagators gives rise to the 
generation of rates where the states on the upper and lower
propagator are changed simultaneously. Rates
describe the evolution of the system into a 
stationary state, i.e. we generate irreversibility
or dissipation during RG.

The RG equations preserve conservation of probability 
and give real expectation values for hermitian observables.
Once the symmetry relations and sum rules, stated in
Eqs.~(\ref{LG-symmetry})-(\ref{sigma-sum})
and (\ref{I-symmetry})-(\ref{sigmaI-symmetry}),
are fulfilled initially, they are not changed by the 
RG-flow. Thus, by applying the same proof as in section
\ref{General approach}, we obtain a normalized 
probability distribution of the dot and a real
expectation value for the current.

Under certain circumstances the RG equations can
be simplified. This will be important for the exact
solution presented in the next section. We decompose
the pair contraction (\ref{gamma-def}) trivially
into two terms
\begin{equation}
\gamma_\mu(t)=\gamma_\mu^\delta(t)+\bar{\gamma}_\mu(t)\,\,,
\label{gamma-decomposition}
\end{equation}
with
\begin{equation}
\gamma_\mu^\delta={1\over 2} \langle [j_{\bar{\mu}}(t),
j_\mu]_{-\sigma}\rangle
\quad,\quad
\bar{\gamma}_\mu={1\over 2} \langle [j_{\bar{\mu}}(t),
j_\mu]_\sigma\rangle\,\,,
\label{delta-bar-def}
\end{equation}
where $\sigma=\pm$ for $\mu=$ bosonic (fermionic), and
$[\cdot,\cdot]_{-\sigma}$ denotes the (anti-)commu\-tator
for $\sigma=\pm$. This
decomposition is useful since in many problems, the
part $\gamma_\mu^\delta(t)$ turns out to be proportional
to a delta function or derivatives of a delta function
(at least for the band width cutoff $D\rightarrow\infty$).
For our special model we get (compare (\ref{gamma-delta}))
\begin{equation}
\gamma_\mu^\delta(t) = {\Gamma_r \over 2}\,\delta(t)\,\,.
\label{delta-part}
\end{equation}
Such a contribution can be included into the initial 
conditions of the RG equations. First, we decompose
$\gamma^{pp'}_{\mu\mu'}(t)=\gamma^{pp'}_{\mu\mu'}(t)^\delta
+\bar{\gamma}^{pp'}_{\mu\mu'}(t)$ as well by inserting 
(\ref{gamma-decomposition}) into (\ref{gamma-1})
\begin{equation}
\gamma^{pp'}_{\bar{\mu}\mu}(t)=
\left[p'\gamma_\mu^\delta(t)+\bar{\gamma}_\mu(t)
\right]\,\left\{
\begin{array}{cl}
1&\mbox{for }\mu=
\mbox{ bosonic}\\
p'&\mbox{for }\mu=
\mbox{ fermionic}
\end{array}
\right.\,\,.
\label{gamma-double-decomposition}
\end{equation}
We have obtained the explicit dependence
on the indices $p$ and $p'$ in this equation. Therefore
we try to include these factors into the definition
of the vertex operators.
Having included the $\gamma^\delta_\mu$-parts into the 
initial conditions, it turns out that closed RG equations can
be found for the quantities
\begin{equation}
(G_\mu)_{ss',\cdot\cdot}\,=\,
\left\{
\begin{array}{cl}
\sum_p (G^p_\mu)_{ss',\cdot\cdot}&\mbox{for }N_s-N_{s'}=
\mbox{ even or }\mu=\mbox{ bosonic}\\
-i\sum_p p(G^p_\mu)_{ss',\cdot\cdot}&\mbox{otherwise }
\end{array}
\right.\,\,,
\label{Gnew-def}
\end{equation}
with $N_s$ being the fermionic particle number of state $s$.
In the same way we define the boundary vertices
$A_\mu$ and $B_\mu$. We have included a factor $-i$ for
the second case in (\ref{Gnew-def}) in order to get the 
same symmetry relations as in (\ref{LG-symmetry})
\begin{equation}
(G_\mu)_{s_1s_1^\prime,s_2s_2^\prime}^*=
(G_{\bar{\mu}})_{s_1^\prime s_1,s_2^\prime s_2}\,\,.
\label{Gnew-symmetry}
\end{equation}
Using the fact that fermionic
vertex operators $(G_\mu)_{ss',s_1s_1'}$ change the
parity of the fermionic particle number difference 
$N_{s_1}-N_{s_1'}\rightarrow N_s-N_{s'}$, we get 
after a lengthy but straightforward calculation the
following RG-equations arising from the 
$\bar{\gamma}_\mu$-parts
\begin{equation}
\fbox{\parbox{12cm}{
\begin{center}
$\begin{array}{rcl}
   {d\over dt_c}\tilde{\Sigma}(z)&=&
   -\int\limits_0^\infty dt\,
   {d\tilde{\gamma}_\mu\over dt_c}(t)
   A{\bar{\mu}}(t)B_\mu
   \\ \\
   {d\over dt_c}L_0&=&
   -i\int\limits_0^\infty dt\,
   {d\tilde{\gamma}_\mu\over dt_c}(t)
   G_{\bar{\mu}}(t)G_\mu\\ \\
   {d\over dt_c}G_\mu&=&
   -\int\limits_0^\infty dt\int\limits_{-\infty}^0 dt'\,
   {d\tilde{\gamma}_{\mu_1}\over dt_c}(t-t')
   \left[\,
   \sigma_{\mu\mu_1}G_{\bar{\mu}_1}(t)
   G_\mu\, - \,G_\mu G_{\bar{\mu}_1}(t)\,\right]\,
   G_{\mu_1}(t')\\ \\
   {d\over dt_c}A_\mu&=&
   -\int\limits_0^\infty dt\int\limits_{-\infty}^0 dt'\,
   {d\tilde{\gamma}_{\mu_1}\over dt_c}(t-t')
   \left[\,
   \sigma_{\mu\mu_1}A_{\bar{\mu}_1}(t)
   G_\mu\, - \,A_\mu G_{\bar{\mu}_1}(t)\,\right]\,
   G_{\mu_1}(t')\\ \\
   {d\over dt_c}B_\mu&=&
   -\int\limits_0^\infty dt\int\limits_{-\infty}^0 dt'\,
   {d\tilde{\gamma}_{\mu_1}\over dt_c}(t-t')
   \,\sigma_{\mu\mu_1} G_{\bar{\mu}_1}(t)
   G_\mu B_{\mu_1}(t')
\end{array}$
\end{center}}}
\label{rg-bar-equations}
\end{equation}

\vspace{3mm}
\noindent
where 
\begin{equation}
\tilde{\gamma}_\mu(t)\,=\,\bar{\gamma}_\mu(t)
\left\{
\begin{array}{cl}
1&\mbox{for }\mu=\mbox{ bosonic}\\
i&\mbox{for }\mu=\mbox{ fermionic}\\
\end{array}
\right.\,\,,
\end{equation}
and
\begin{equation}
\sigma_{\mu\mu'}\,=\,\left\{
\begin{array}{cl}
1&\mbox{for }\mu\mbox{ or }\mu'=
\mbox{ bosonic}\\
-1&\mbox{for }\mu\mbox{ and }\mu'=
\mbox{ fermionic}\\
\end{array}
\right.\,\,.
\end{equation}
We see that the new RG equations are more compact but
we note that they can not be applied to all models.
Sometimes, like e.g. in spin boson models, it happens
that the part $\gamma^\delta_\mu(t)$ can not be
expressed by the derivative of a delta function and,
consequently, can not be incorporated into the
initial conditions. In this case, one should use
the original and generally valid RG-equations
(\ref{rg-equations}).

\section{Exact solution of the RG equations}
\label{Exact solution of the RG equations}

In this section we demonstate that the RG equations
can be solved exactly for the special model 
(\ref{Hres})-(\ref{Htun}) of a single non-degenerate dot state.
It turns out that the result for the kernels
agrees with that of section \ref{Exact solution}.
We conclude that the RG-approach solves the present
model exactly.

We choose the sharp cutoff-function $F(x)=\theta(x-1)$
and get from the RG-equation (\ref{rg-equations}) for
a matrix element of the kernel
\begin{equation}
\tilde{\Sigma}(z)_{ss,\bar{s}\bar{s}}\,=\,
\sum_{pp'\mu s'}\int\limits_0^\infty dt_c\,\,
\gamma^{pp'}_{\bar{\mu}\mu}(t_c)\,e^{izt_c}\,
(A^p_{\bar{\mu}})_{ss,s'\bar{s}'}\,
e^{-i\alpha_{s'}t_c}\,
(B^{p'}_\mu)_{s'\bar{s}',\bar{s}\bar{s}}\,\,.
\label{sigma-calculation}
\end{equation}
Here, due to particle number conservation, we have used 
the fact that the only nonvanishing matrix elements of $L_0$ are 
$\alpha_s=(L_0)_{s\bar{s},s\bar{s}}$ and $(L_0)_{ss,s's'}$.
To evaluate this equation we need the $t_c$ dependence
of $\alpha_s$ and the boundary vertex operators. To obtain
this it is convenient to include the part 
$\gamma^\delta_\mu$ into the initial conditions. 
The latter follow from inserting the part
$\gamma^{pp'}_{\bar{\mu}\mu}(t_c)^\delta$ into the
RG-equations (\ref{rg-equations}), and integrating.
We denote the initial value for $L_0$ by $L_0^\delta$.
The vertex operators do not obtain any initial 
renormalization from the $\delta$-function parts since
there is no phase space for the time integrals in
(\ref{rg-equations}). Taking the part 
$\gamma^{pp'}_{\bar{\mu}\mu}(t_c)^\delta$
from the first term of (\ref{gamma-double-decomposition})
together with (\ref{delta-part}), and inserting it in
the RG-equation (\ref{rg-equations}) for $L_0$, we find
\begin{equation}
({d\over dt_c}) \alpha_s^\delta = i{\Gamma\over 2}
\delta(t_c) pp'(G^p_{\bar{\mu}}G^{p'}_\mu)_
{s\bar{s},s\bar{s}} = -i\Gamma\delta(t_c)\,\,.
\end{equation}
This gives
\begin{equation}
\alpha_0^\delta = -\epsilon-i{\Gamma\over 2}
\quad,\quad
\alpha_1^\delta = \epsilon-i{\Gamma\over 2}\,\,,
\label{alpha-value}
\end{equation}
where the part with the single particle energy $\epsilon$
stems from the initial value without any renormalization.

As we will show in the following we find no further 
renormalization from the $\tilde{\gamma}_\mu$-part, i.e.
using the simplified set of RG-equations 
(\ref{rg-bar-equations}), we get
\begin{equation}
{d\over dt_c}\alpha_s = 
{d\over dt_c}(A_\mu)_{ss,s'\bar{s}'} =
{d\over dt_c}(B_\mu)_{s'\bar{s}',\bar{s}\bar{s}} =
{d\over dt_c} G_\mu = 0\,\,.
\label{ren=null}
\end{equation}
This means that the renormalization of the kernel
due to the $\tilde{\gamma}_\mu$-part can be calculated
with unrenormalized vertex operators and using the result
(\ref{alpha-value}) for $\alpha_s$. Since, trivially,
the same applies to the $\gamma^\delta_\mu$-part,
we can directly evaluate (\ref{sigma-calculation})
in this way and get the result (\ref{exact-solution}) of 
section \ref{Exact solution}.

What remains to be shown is Eq.~(\ref{ren=null}). We
first note the following properties of the matrix
elements of the vertex operator, which follow 
from the definition (\ref{Gnew-def})
\begin{equation}
\sum_s(G_\mu)_{ss,s'\bar{s}'}=0
\quad,\quad
(G_\mu)_{s\bar{s},s's'}=
(G_\mu)_{s\bar{s},\bar{s}'\bar{s}'}\,\,.
\end{equation}
Furthermore, using (\ref{LABG-sum}), we get
\begin{equation}
\sum_s (e^{-iL_0t})_{ss,\cdot\cdot}=1\,\,.
\end{equation}
We apply these properties to the following expression
\begin{eqnarray}
\left[G_\mu(t)G_{\mu'}(t')\right]_{s\bar{s},s'\bar{s}'}&=&
\nonumber\\
&&\hspace{-2cm}
= e^{i\alpha_s t}e^{-i\alpha_{s'}t'}
\sum_{s_1s_2}(G_\mu)_{s\bar{s},s_1s_1}
(e^{-iL_0(t-t')})_{s_1s_1,s_2s_2}
(G_{\mu'})_{s_2s_2,s'\bar{s}'}\nonumber\\
&&\hspace{-2cm}
= e^{i\alpha_s t}e^{-i\alpha_{s'}t'}
(G_\mu)_{s\bar{s},s_1s_1}
\sum_{s_2}(G_{\mu'})_{s_2s_2,s'\bar{s}'}
\quad =0 \,\,.
\end{eqnarray}
Using this result after taking matrix elements of 
the RG-equations (\ref{rg-bar-equations}), we find
immediately (\ref{ren=null}).

We conclude that after having set up the general RG-equations
(\ref{rg-equations}) and (\ref{rg-bar-equations}),
the exact solution of the present model can be found 
analytically in a straightforward way. There is no need
to consider diagrammatic details as in section
\ref{Exact solution}. Furthermore, we see that the
effect of the reservoirs is simply a broadening of
the local state of the dot by $\Gamma/2$, see
(\ref{alpha-value}). Otherwise the evaluation of
the kernel is the same as in lowest order perturbation
theory in $\Gamma$.

\section{Summary and outlook}
\label{Summary and outlook}

In this article we have presented a new viewpoint
to analyse nonperturbative aspects of nonequilibrium
systems. We considered a small system coupled linearly
to several baths. Nonperturbative means that the
coupling between system and bath is so strong that
quantum fluctuations induce broadening of the states
in the system together with possible renormalization
of energy levels and the coupling to the environment.
In macroscopic systems, such effects are negligible
since the interaction with the environment is a 
surface effect. In contrast,
our goal is to describe mesoscopic systems like
quantum dots, magnetic nanoparticles or chemical
molecules coupled to particle or heat reservoirs.
Usually such systems are treated within
perturbation theory \cite{blum,fick,gar}. This means 
that the kernel $\Sigma(t)$ of the kinetic equation
is calculated in lowest order perturbation theory 
in the coupling, the so-called golden rule
or Pauli-Master equation approach. 
Our aim here was to find a systematic way to consider
self-consistently an infinite series of higher-order
contributions to the kernel. This is a nontrivial
task, especially in nonequilibrium where a real-time
formalism on a Keldysh contour has to be used.
Renormalization effects known from
equilibrium theories have to be incorporated 
consistently within a kinetic equation.
Our point of view relies on renormalization group
ideas. We try to integrate out all energy scales
of the bath in infinitesimal steps. Each step is
interpreted as a change of various quantities,
like broadening and renormalization of energy levels, 
and generation of rates. For formal reasons we have set
up this procedure in real-time space. During this
procedure we keep the kernel of the kinetic equation
invariant and generate non-Hamiltonian dynamics to
describe the physics of dissipation. The final
RG-equations are presented in (\ref{rg-equations}),
and, alternatively, in (\ref{rg-bar-equations}). 
Solving them, provides the complete description
of the time-evolution of the reduced density matrix
of the system and arbitrary observables being linear
in the field operators of the bath. This includes
the consideration of an initally out of equilibrium
state as well as the description of stationary
nonequilibrium situations. We have solved the equations
exactly for the special case of a quantum dot with
one state coupled to two particle reservoirs. The
solution turned out to be identical to the exact
one.

The reader might argue that this is a trivial result
since the model under consideration is very simple.
However, as shown in a recent paper, the same 
RG equations provide not only the exact solution
for a noninteracting quantum dot with one state, but 
gives a good solution also
for the spin-degenerate case including a finite
on-site Coulomb interaction $U$ \cite{RG-anderson},
the so-called Anderson impurity model in 
nonequilibrium. Here, not only energy broadening
but also energy and coupling constant 
renormalizations are important. In the mixed-valence 
and empty-orbital
regime, i.e. for level positions near or above the
Fermi level of the reservoirs, it was shown that
the linear conductance and the average occupation
in equilibrium agrees perfectly with Friedel sum
rules, Bethe ansatz, and numerical renormalization
group methods within 2-3\%. This means that the
RG provides a good solution for the whole range
from $U=0$ to $U=\infty$. This is a surprising
result since usually methods designed for strong
interaction do not work well for weak interaction
and vice versa. Therefore, the fact that the RG
gives the exact result for the noninteracting
case is not at all a trivial result. In fact,
within the slave boson technique \cite{barnes}, 
which is a well-known and well-established method in the 
theory of strongly correlated Fermi systems, it 
is very complicated to find the exact solution 
for $U=0$ \cite{barnes}.

Another example where the RG-method has been applied
is the study of transport through the metallic 
single-electron transistor \cite{kuc-etal}. For this
case, the RG-equations have been solved in sixth-order 
perturbation theory in the coupling. Surprisingly,
it turned out that the solution agrees very well
with exact perturbation theory in the same order.
The reason why the RG-equations have not been solved
in all orders is that the result was not finite
for the band width cutoff $D\rightarrow\infty$. Only
if certain parts of double-vertex corrections
were included, the solution turned out to be
cutoff-independent. A more detailed study of the
$D$-dependence and the solution for higher orders
is currently under way.

The RG-equations on the level of propagator and
single-vertex renormalization are pure differential
equations and, consequently, can be solved numerically
very quickly. Typical response times range from seconds
to a few minutes for one set of parameters, even on
a usual PC maschine. There are still some problems
with the asymptotic solution of the RG equations
for $t_c\rightarrow\infty$. First, it is not known 
rigorously wether a stationary solution exists at all.
Secondly, the numerical solution is plagued by
oscillating functions, typical for real-time
problems. However, for the problems studied so far,
and provided the coupling is not too strong, there
is a stationary numerical solution over a sufficiently
long period in $t_c$. This applies at least to the
physical quantities under consideration, like the
probability distribution and the current.

Another reason for the efficiency of the present method 
is the possibility to calculate physical observables 
directly without the need of correlation functions
like in linear response theory. Furthermore, if desired, 
correlation functions can also be studied
with the RG-method. In this case, there are additional
RG-equations to describe the renormalizations of the
external vertices defining the correlation function.
These equations together with
explicit solutions will be published in forthcoming
works.

The method is also applicable
to the study of the ground state energy since the
RG-flow on the single forward propagator provides the
S-matrix. This idea has been applied to the single-electron
box \cite{seb}, coupled metallic islands
\cite{poh-etal}, and the one-dimensional Polaron problem
\cite{keil-hs}. In the first two cases very good
results have been obtained, even comparable to very
time-consuming QMC-simulations. For the 1d-Polaron
problem, the results were at least satisfactory for small
coupling. Here, problems occured since the correlation
function of the bath does not decay for long times and
undamped modes occur which make the numerical analysis
of the asymptotic solution very difficult.

Finally, we remark that a challenge for future research
is the consideration of higher-order vertex corrections.
Our RG-scheme provides a systematic treatment
for setting up RG-equations for all kinds of multiple 
vertices. However, even on the level of double-vertices,
the number of terms increases considerably and the
RG-equations become integral-differential equations.
The reason is the retarded nature of double-vertices.
Therefore it is necessary to find physical arguments
to select the most important terms or to improve the
numerical efficiency by neglecting the retardation in
a convenient way. Whereas single-vertices describe
basically charge fluctuations (in case of coupling to 
particle reservoirs), double-vertices describe physical 
processes via virtual intermediate states where the 
local system can change its state without 
changing its particle number. This is e.g. important 
for the study of spin fluctuations in local impurities or
quantum dots, leading to the Kondo effect
\cite{kondo-exp,hewson}. Although such
processes are perturbatively included in the RG-equations
set up in this article, it is important to study
them fully self-consistently by considering corresponding
RG-equations for double-vertex terms on a Keldysh
contour.

\section{Acknowledgements}

Useful discussions are acknowledged with T. Costi, 
J.v. Delft, H. Grabert, G. Sch\"on, K. Sch\"onhammer, 
and P. W\"olfle. I owe special thanks to J. K\"onig,
T. Pohjola, and M. Keil for very fruitful collaborations 
on applying the RG-method to various systems.
This article was supported by the Swiss 
National Science Foundation (H.S.) and by the ''Deutsche 
Forschungsgemeinschaft'' as part of ''SFB 195''.

\appendix

\section*{Appendix}
Here we prove Eqs.(\ref{prop1})-(\ref{prop3}). The
first one follows from
\begin{equation}
(G^+_\mu
  \begin{picture}(-10,11)
    \put(-10,8){\line(0,1){3}}
    \put(-10,11){\line(1,0){15}}
  \end{picture}
  \begin{picture}(10,11)
  \end{picture}
)_{ss,\bar{s}s}\,+\,
(G^-_\mu
  \begin{picture}(-10,11)
    \put(-10,8){\line(0,1){3}}
    \put(-10,11){\line(1,0){15}}
  \end{picture}
  \begin{picture}(10,11)
  \end{picture}
)_{\bar{s}\bar{s},\bar{s}s}\,=\,0\,\,,
\end{equation}
which can easily be seen by using the definitions
(\ref{def-gmatrix1}) and (\ref{def-gmatrix2}). The reservoir 
contraction is the same in both terms since $G^p_\mu$ is the 
vertex at later time, compare (\ref{gamma-2}). The same proof
can be used for $I^p_\mu$ by using the definition
(\ref{def-curmatrix}). Since there is no change
of sign of $I^p_\mu$ if $p$ is changed to $\bar{p}$,
we have to add a factor $p$ under the sum in
(\ref{prop1}).

To show the second property (\ref{prop2}), we again
use the definitions (\ref{def-gmatrix1}) and 
(\ref{def-gmatrix2}), and find
\begin{equation}
(G^p_\mu
  \begin{picture}(-10,11)
    \put(-10,8){\line(0,1){9}}
    \put(-10,17){\line(1,0){30}}
  \end{picture}
  \begin{picture}(10,11)
  \end{picture}
G^{p'}_{\mu'}
  \begin{picture}(-11,11)
    \put(-11,8){\line(0,1){5}}
    \put(-11,13){\line(1,0){15}}
  \end{picture}
  \begin{picture}(11,11)
  \end{picture}
)_{s\bar{s},\cdot\cdot}\,+\,
(G^{\bar{p}}_\mu
  \begin{picture}(-8,11)
    \put(-8,8){\line(0,1){9}}
    \put(-8,17){\line(1,0){28}}
  \end{picture}
  \begin{picture}(8,11)
  \end{picture}
G^{\bar{p}'}_{\mu'}
  \begin{picture}(-11,11)
    \put(-11,8){\line(0,1){5}}
    \put(-11,13){\line(1,0){15}}
  \end{picture}
  \begin{picture}(11,11)
  \end{picture}
)_{s\bar{s},\cdot\cdot}\,=\,0\,\,.
\label{proof-2}
\end{equation}
Both reservoir contraction are the same in both
terms since both vertices are at later time.
However, the reader can convince himself very
easily that there is an additional relative sign
between the two terms of (\ref{proof-2}) due to
the interchange of reservoir Fermi operators.

The third property (\ref{prop3}) follows from
\begin{equation}
\lim_{D\rightarrow\infty}\{(G^p_\mu
  \begin{picture}(-8,11)
    \put(-8,8){\line(0,1){6}}
    \put(-8,14){\line(1,0){25}}
  \end{picture}
  \begin{picture}(8,11)
  \end{picture}
G^{p'}_{\mu'}
  \begin{picture}(-11,11)
    \put(-11,8){\line(0,1){3}}
    \put(-11,11){\line(-1,0){20}}
  \end{picture}
  \begin{picture}(11,11)
  \end{picture}
)_{s\bar{s},\cdot\cdot}\,+\,
(G^{\bar{p}}_\mu
  \begin{picture}(-8,11)
    \put(-8,8){\line(0,1){6}}
    \put(-8,14){\line(1,0){25}}
  \end{picture}
  \begin{picture}(8,11)
  \end{picture}
G^{\bar{p}'}_{\mu'}
  \begin{picture}(-11,11)
    \put(-11,8){\line(0,1){3}}
    \put(-11,11){\line(-1,0){20}}
  \end{picture}
  \begin{picture}(11,11)
  \end{picture}
)_{s\bar{s},\cdot\cdot}\}\,=\,0\,\,.
\end{equation}
Here, by comparing the two terms, the contraction 
associated with the second
vertex changes from $\gamma^\eta_r$ to 
$\gamma^{-\eta}_r$. Otherwise
the two terms are the same. However, the sum
$\gamma^\eta_r(t)+\gamma^{-\eta}_r(t)$ gives a 
$\delta(t)$-function in the limit 
$D\rightarrow\infty$, see (\ref{gamma-delta}).
This means that there is no phase space in time
for the first vertex and the sum is zero.

\end{document}